\documentclass[lettersize,journal]{IEEEtran}
\usepackage{amsmath,amsfonts,amssymb}
\usepackage{algorithmic}
\usepackage{algorithm}
\usepackage{array}
\usepackage[caption=false,font=normalsize,labelfont=sf,textfont=sf]{subfig}
\usepackage{textcomp}
\usepackage{stfloats}
\usepackage{url}
\usepackage{tabularx}
\usepackage{verbatim}
\usepackage{graphicx}
\usepackage{cite}
\usepackage{comment}
\usepackage[table]{xcolor}
\usepackage[hidelinks]{hyperref}
\usepackage{booktabs}
\usepackage{multirow}
\usepackage{array}
\usepackage{comment}
\usepackage{bbm}

\usepackage{makecell}

\usepackage{pifont}
\usepackage{amssymb}
\usepackage{colortbl}
\usepackage{bm}

\usepackage[nolist,nohyperlinks]{acronym}
\begin{acronym}
    \acro{mmWave}{millimeter wave}
    \acro{Mbps}{megabits per second}
    \acro{Mbps}{gigabits per second}
    \acro{DAC}{digital to analog converter}
    \acro{ADC}{analog to digital converter}
    \acro{BER}{bit error rate}
    \acro{RRC}{root-raised cosine}
    \acro{digital up converter}{digital up-converter}
    \acro{digital down converter}{digital down-converter}
    \acro{TX}{transmitter}
    \acro{RX}{receiver}
    \acro{MSPS}{mega samples per second}
    \acro{MSPS}{giga samples per second}
    \acro{RFSoC}{radio frequency system-on-chip}
    \acro{AFE}{analog front end}
    \acro{OFDM}{orthogonal frequency division multiplexing}
    \acro{PS}{processing system}
    \acro{PL}{programmable logic}
    \acro{BER}{bit error rate}
    \acro{RF}{radio frequency}
\end{acronym}

\newcolumntype{C}[1]{>{\centering\arraybackslash}p{#1}}

\begin{document}

\title{Prototype and Real-Time Over-The-Air Evaluation of Millimeter Wave Integrated Sensing and Communication Transceiver at Edge}

\author{Jai Mangal, Antriksh Choudhary, Ahilan R, Sumit Darak, Utkarsh Tripathi, Shobha Sundar Ram.
\thanks{This paper is an expanded version from the 2026 IEEE International MTT Symposia (IMS): RF Technology \& Techniques (RFTT) and RF Systems \& Applications (RFSA) Symposia, Boston, Massachusetts, USA, 7-12 June, 2026.}
\thanks{The first five authors are with the Indraprastha Institute of Information Technology, Delhi, New Delhi 110020 India. E-mail: \{jaim, antriksh23106, ahilan24170, sumit, utkarsh25154\}@iiitd.ac.in. Last author is with University College London, United Kingdom, Email: \{s.ram@ucl.ac.uk\}}}

\markboth{Journal of \LaTeX\ Class Files,~Vol.~14, No.~8, August~2021}%
{Shell \MakeLowercase{\textit{et al.}}: A Sample Article Using IEEEtran.cls for IEEE Journals}

\maketitle

\begin{abstract}
This paper presents the design and experimental validation of an end-to-end millimeter-wave integrated sensing and communications testbed at 28~GHz for next-generation vehicular networks. The prototype uses the Xilinx radio frequency system-on-chip 4x2 as the digital frontend and the Sivers EVK02004 and EVK02001 as the analog frontends for the base station and mobile user, respectively. Radar and communication signal processing is implemented entirely in real time on the edge via a hardware-software co-design approach, achieving acceleration factors of 7.10$\times$ and 52.10$\times$, respectively, over the software baseline.  We characterize nondeterministic pipeline jitter in the digital frontend that induces a 2.5~m ambiguity range, which cannot be removed by offline calibration. A custom synchronization core is designed to compensate for it in real time, bounding the range accuracy by the Fourier limit. Further, over-the-air measurements yield root-mean-square errors of 0.11 m in range and 2.69$^\circ$ in azimuth, resolving up to 5 targets in a real radio environment. Subsequently, sensing-assisted beam-aligned directional communication achieves an error vector magnitude of $-12$~dB, a bit error rate of $10^{-5}$, and a throughput of 0.20~Gbps using 16-QAM modulation over a 153.6~MHz bandwidth. We further quantify how radar beam misalignment degrades these metrics and demonstrate ISAC for sensing-assisted communication, in which the degradation in communication metrics reinitiates the sensing stage that restores beam alignment as the user moves.
\end{abstract}


\section{Introduction}
Next-generation transportation systems will simultaneously require high-resolution environmental sensing and multi-gigabit data communication at ultralow latency to improve road safety and passenger comfort~\cite{survey_zhang_2023}. Existing sub-6 GHz vehicular communication protocols cannot support these requirements due to limited bandwidth. Solutions are therefore being actively researched in the millimeter wave (mmWave) frequency range, which offers less spectrum congestion, wider bandwidths, and smaller component sizes~\cite{ieee_nitsche_2014}. However, path loss at these frequencies is significant due to high atmospheric absorption, demanding narrow directional links. As a result, in high-mobility automotive environments, the base station (BS) must rapidly detect and localize the mobile users (MU) to establish a directional communication link. Existing mmWave communication protocols, such as the IEEE 802.11ad standard, address this step through a lengthy beam alignment procedure using additional beam-training fields within the communication packet. However, this procedure introduces significant communication overhead and deteriorates latency. Alternatively, researchers have recently proposed using common waveforms, hardware, and spectral resources for both sensing and localizing the MU, and for subsequent communication, through BS implemented with integrated sensing and communication (ISAC) transceivers~\cite{ieee_kumari_2017,duggal2020Doppler}. The advantages of this approach are threefold: (1) hardware and spectrum are reused, reducing hardware cost and spectrum congestion \cite{SnehTVT2024}; (2) synchronization and interference between sensing and communication are managed by implementing both on a common platform; and (3) sensing can be exploited to enhance communication performance by enabling accurate estimation of MU location (for beamforming)~\cite{Mangal2025MAES}, distance (for adaptive transmission parameter selection)~\cite{Mangal2025URSI}, and velocity (for efficient handover management)~\cite{Sneh2023LCOMM}.

Several works have studied the theoretical performance and limitations of ISAC systems in automotive environments~\cite{ieee_kumari_2017,tewari2024reconfigurable, Gan2024JSTAC, Duan2025IOT, Su2025IOT}. However, hardware testbeds capable of evaluating real-time performance in real radio environments are required for studying joint radar-communication metrics and the mutual effects of their interplay. Currently, commercial mmWave radar platforms are available with radio frequency system on chip (RFSoC) modules featuring wideband frequency and phase-modulated continuous-wave waveforms and multiple-input multiple-output (MIMO) antenna configurations~\cite{Wadaskar2025CICC, Fhager2019LSENS, Jiang2025TMTT}. Similarly, several commercial software-defined radio (SDR) platforms have demonstrated massive MIMO at mmWave frequencies, achieving high communication throughput~\cite{Zetterberg2015ACCESS, Ng2025TMTT, Zhang2019ACCESS}. However, these radar and communication platforms remain separate systems, and commercial off-the-shelf ISAC hardware for radar-enhanced communication is not yet available.

Early ISAC prototypes at 28~GHz~\cite{3d_guan_2021, mmwave_ozkaptan_2024}, 60~GHz~\cite{openmili_demo_2016, mcube_mobicom_2020}, and 73~GHz~\cite{jcr70_kumari_2021} processed radar and communication signals offline on a personal computer and hence did not support low-latency signal processing. A recent work in ~\cite{zhou2023iccc} addressed this limitation by demonstrating real-time edge processing. However, distinct challenges involving hardware-level synchronization that introduce ambiguity in the range estimation have neither been characterized nor compensated in real-time on existing platforms. Further, integrating radar sensing output into the communication chain in real time to fully realize ISAC operation remains an open challenge, and the impact of radar hardware impairments on communication performance metrics such as bit error rate (BER) and error vector magnitude (EVM) has not been experimentally studied.

\begin{figure}
    \centering
    \includegraphics[width=0.5\textwidth, trim= 0cm 4.3cm 0cm 0cm, clip]{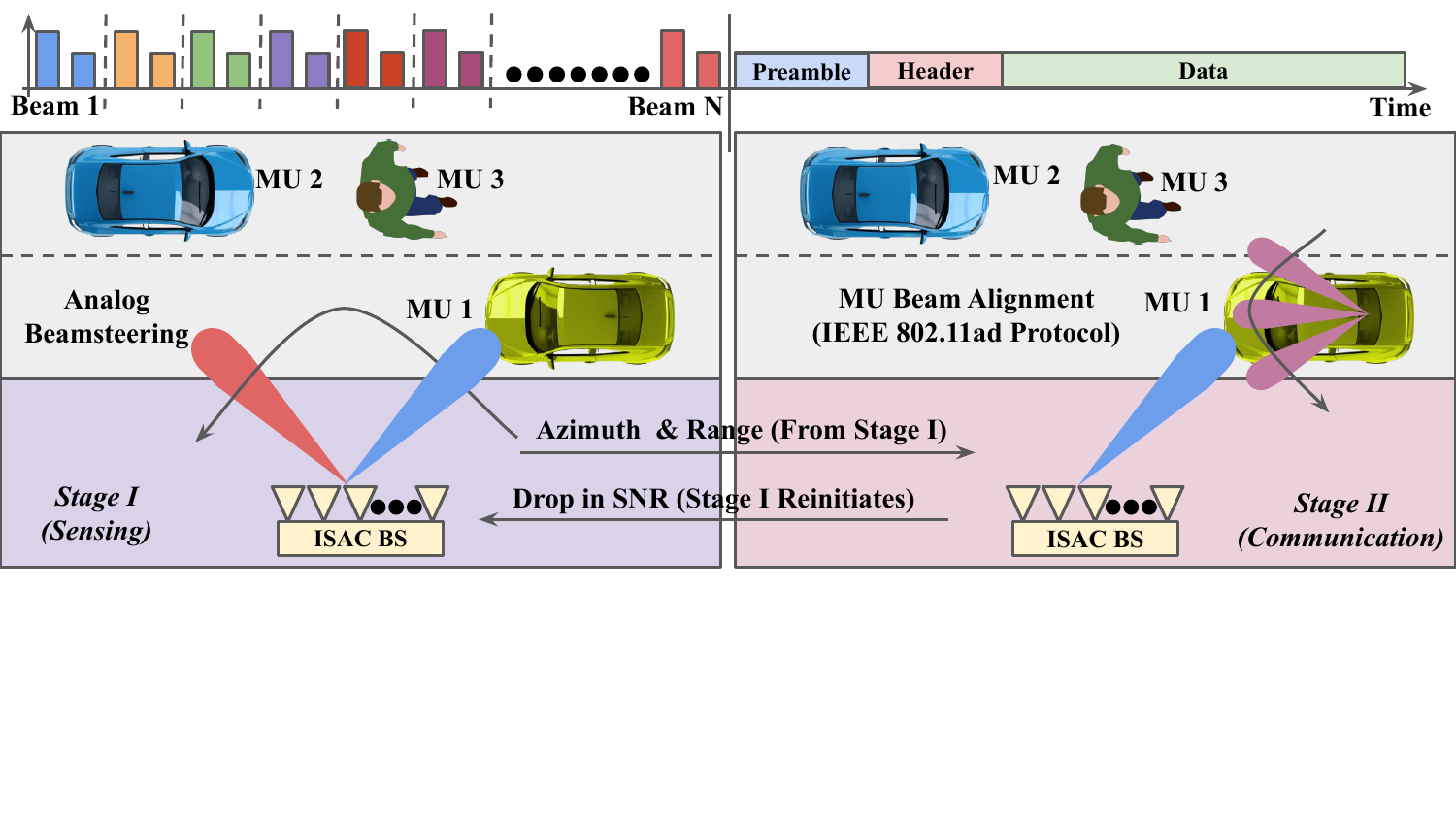}
    \caption{Proposed TDM-based two-stage ISAC protocol for V2X environment.}
    \label{fig:ISAC_System}
\end{figure}

This work extends \cite{Mangal2026rfsa} by addressing these gaps by presenting an end-to-end mmWave ISAC testbed operating at 28 GHz. The testbed is built using the Xilinx RFSoC 4x2 as the digital frontend (DFE) and the Sivers EVK02004 as the analog frontend (AFE), with real-time waveform generation and signal processing on edge. The IEEE 802.11ad communication waveform is used, chosen for its preamble structure that embeds Golay complementary sequences. These sequences are exploited for radar processing due to their ideal autocorrelation properties. The system operates as a full-duplex ISAC platform in a time-division-multiplexing (TDM) format, where radar sensing and communication alternate within a structured two-stage protocol, as illustrated in Fig.~\ref{fig:ISAC_System}. Therefore, the duration of real-time radar signal processing for searching the channel space for MU must be minimized to enhance subsequent communication service with the detected MU. Second, since the radar identifies and localizes the MU, the range and angular estimation accuracy achieved by the radar directly affects the BER and EVM of subsequent communication. Thirdly, radar hardware impairments, such as non-deterministic jitter in the signal pipeline, degrade coherence between the ISAC BS transmitter (TX) and BS receiver (RX), thereby impacting range estimation accuracy and, ultimately, the detection of the MU. The novelty of our research presented in this work is the qualitative and quantitative analysis of the inter-dependence between radar and communications, specifically, how hardware-limited radar metrics directly govern subsequent communication performance.

The primary contributions are as follows: (1) We present a fully integrated ISAC platform combining the DFE and AFE on the Xilinx RFSoC 4x2 and Sivers EVK02004 platforms, respectively. We have performed hardware-software co-design to implement custom IP cores within the programmable logic (PL) for both the radar and communication processing chains. The radar and communication signal processing  achieve acceleration factors of 7.10$\times$ and 52.10$\times$, respectively, over the processing system (PS) baseline.
Together, these enable real-time dual-functional operation in a resource-efficient form factor suitable for vehicular deployment. (2) The DFE is characterized by a non-deterministic pipeline jitter that introduces a range ambiguity of 2.5 m, five times the range resolution. This impairment is invisible in simulation and only manifests under full over-the-air operation, and cannot be removed by offline calibration.
We implement a custom synchronization IP core within the RFSoC PL to measure and compensate for jitter in real time, ensuring range estimation accuracy remains bounded by the Fourier limit. (3) We conduct comprehensive over-the-air measurements to both independently and jointly evaluate the radar and communication subsystems, resulting in root mean square error (RMSE) of 0.11 m and 2.69$^\circ$ for radar range and angle estimation, respectively, and the system demonstrates a capability of resolving five targets.
Similarly, the communication system's EVM and BER are evaluated across different bandwidths, modulation and coding schemes (MCS), and transmission distances. The proposed testbed achieves an EVM of -12 dB and a BER of $10^{-5}$, with a $\frac{1}{2}$ code rate, 16-QAM modulation, and 153.6 MHz of bandwidth.
Finally, we evaluated how radar-based beam misalignment degrades the communication BER and EVM. (4) We have experimentally demonstrated the real-time closed-loop operation of radar-enhanced communication operating in a two-stage TDM manner. In the first stage, the radar extracts the angle and range of the MU, which are used by the communication subsystem to perform directional communication during the second stage. When the target/MU moves outside the BS mainlobe, the communication metrics deteriorate, triggering the radar sensing stage  to recommence to acquire the new location of the MU.

The remainder of this paper is organized as follows. Section~\ref{sec:system_model} presents the system model and waveform design, including the radar and communication signal models. Section~\ref{sec:testbed} presents the proposed ISAC testbed hardware and hardware synchronization and calibration procedures, along with the receiver signal processing pipeline, detailing the PL-accelerated architectures. Section~\ref{sec:results} presents the experimental results and performance analysis of the proposed ISAC system. Section~\ref{sec:conclusion} concludes the paper with a discussion of future research directions. The notation conventions used throughout this paper are defined below.

\emph{Notation:} Bold lowercase and uppercase letters represent vectors and matrices, respectively, whereas regular characters are used for scalar variables. The superscripts $(\cdot)^T$, $(\cdot)^H$, and $\circledast$ denote the transpose, the complex conjugate transpose, and convolution operations, respectively. $\hat{(\cdot)}$ and $\arg(\cdot)$ denote an estimated quantity and the phase angle of a complex number in the range $(-\pi, \pi]$. $\mathbb{R}$, $\mathbb{B}$, and $\mathbb{C}$ correspond to real, binary, and complex-valued quantities, respectively. The operators $\mathcal{F}\{\cdot\}$ and $\mathcal{F}^{-1}\{\cdot\}$ represent the forward and inverse fast Fourier transform (FFT), respectively. The notation $\mathcal{O}(\cdot)$ denotes the computational complexity order.

\section{System Model}
\label{sec:system_model}
In this section, we present the complete system model of the proposed ISAC testbed. The ISAC functionality is implemented in two stages in a TDM manner, as shown in Fig.~\ref {fig:ISAC_System}. Here, we assume that the radar and communication functionalities share a common channel, and the MU of the communication is a mobile target to be detected by the radar. Hence, both terms are used interchangeably throughout this paper. In Stage I, the BS radar detects and localizes the MU in the range-azimuth space. Subsequently, in Stage II, this information is used to perform beam-aligned directional communication. We implement an IEEE 802.11ad-based ISAC waveform, with each packet comprising a preamble, a header, and a data payload. The optional beam-training fields used in the standard IEEE 802.11ad packet are omitted in the ISAC waveform, since beam alignment is achieved through radar sensing.
The process is detailed in the following subsections.

\subsection{Radar Signal Model and Processing}
The radar waveform is based on the preamble of the IEEE 802.11ad waveform. Length-$N$ complementary Golay sequences, denoted as $\mathbf{g}_d[n],\ d \in \{a,b\}$, with a sampling interval of $T_s$, are generated on the DFE of the RFSoC 4x2. These sequences are chosen because the sum of their individual autocorrelations yields a perfect impulse, with zero sidelobes. Typically, the Doppler shift introduced by fast moving targets can degrade the perfect autocorrelation property of complementary sequences. To address this limitation, we consider the Doppler-resilient complementary sequences for the radar waveform as discussed in \cite{duggal2020Doppler}. The baseband sequences are digitally modulated to an intermediate frequency (IF) $f_{\text{IF}}$, generated by a numerically controlled oscillator. The discrete-time ($n$) baseband signal is then converted to continuous-time ($t$) through a digital-to-analog converter (DAC) and shaped with a transmit pulse shaping filter $\mathbf{w}(t)$, as follows:
\begin{equation}
\mathbf{g}_d(t) = \sum_{n=0}^{N-1} \mathbf{g}_d[n]\, e^{j 2\pi f_{\mathrm{IF}} nT_s}\, \mathbf{w}(t - nT_s)
\label{eq:golay_tx}
\end{equation}
Both the complementary Golay sequences are transmitted with a pulse repetition interval, $T_{\mathrm{PRI}}$. The radar waveform is a pulse-modulated continuous-wave, single-carrier waveform, in which $\mathbf{g}_a$ and $\mathbf{g}_b$ are transmitted in two consecutive PRIs of $3.33~\mu$s with a duty cycle of $50\%$. The signal is then upconverted in the AFE to the mmWave carrier frequency $f_c$ using a local oscillator. Then, spatial sensing is performed via sequential analog beamsteering using a uniform linear array (ULA) of $N_t$ and $N_r$ elements in the BS-TX and BS-RX, respectively. In both cases, the elements are spaced with half-wavelength apart and $\mathbf{p}(\phi_i) \in \mathbb{C}^{N_t \times 1}$ $\mathbf{q}(\phi_i) \in \mathbb{C}^{N_r \times 1}$ are the transmitting and receiving weight vectors, respectively, for each steering angle $\phi_i$ within the angular space $\Omega$. We assume both the ULAs are physically close, parallel, and synchronized with each other, allowing for a full-duplex monostatic radar operation such that they scan the beams together across $\Omega$.
We assume an $m^{th}$ moving target, located at range $r_m$, angle $\phi_m$, moving with radial velocity $v_m$, with amplitude $a_m$ that includes two-way path loss from the BS-TX to the target and back to BS-RX.
The spatial gain between the BS-TX, target, and BS-RX is
\begin{equation}
G(\phi_m, \phi_i) = \mathbf{q}^H[\phi_i]\mathbf{v}(\phi_m)\mathbf{u}^H(\phi_m)\mathbf{p}[\phi_i],
\end{equation}
where $\mathbf{u}$ is the transmitting steeering vector
\begin{equation}
\mathbf{u}(\phi_i) = \left[1,\, e^{j\pi\sin\phi_m},\, \ldots,\, e^{j\pi(N_t-1)\sin\phi_m}\right]^T,
\label{eq:tx_weight}
\end{equation}
and $\mathbf{v}$ is the corresponding steering vector at the receiver ULA.
$G(\phi_m,\phi_i)$ is maximized when the beamsteering angle matches the target angle. The time duration corresponding to each beam is $2 \times T_{PRI}$ as shown in the Fig.~\ref{fig:ISAC_System}. The target introduces a time delay of $t_m = \frac{2r_m}{c}$ where $c$ is the speed of light, and a Doppler frequency shift $f_m = \frac{2v_mf_c}{c}$. If we assume that all $M$ targets are within the field of view of the BS and are within the maximum unambiguous range, then the received signal, for every $i^{th}$ beam, after downconversion and digitization, is
\begin{equation}
\small
\mathbf{y}_{d}[n, \phi_i] =
\sum_{m=1}^M a_m G(\phi_m,\phi_i)\mathbf{g}_d[n-n_m]
e^{-j2\pi f_m nT_s}
+\mathbf{z}_d
\label{eq:radar_tx}
\end{equation}
where $\mathbf{z}_d$ is the additive white Gaussian noise. In this discussion, we assume slow-moving/static targets and rapid electronic beam scanning. Hence, the Doppler phase shift is negligible and is not processed. In other words, we do not consider slow time processing in this work.  Instead, radar signal processing is performed to localize the target in range and azimuth.

Matched filtering in the time domain is performed by convolving the received signal with the time-reversed conjugate of the respective transmit Golay sequences. The correlated outputs of both Golay sequences are then summed to exploit their ideal complementary autocorrelation properties, resulting in complete range sidelobe cancellation, as shown:
\begin{equation}
\mathbf{\chi}[n,\phi_i] = \mathbf{y}_{a}[n,\phi_i] \circledast \mathbf{g}_{a}^H[-n] + \mathbf{y}_{b}[n,\phi_i] \circledast \mathbf{g}_{b}^H[-n]
\label{eq:xcorr}
\end{equation}
Target detection and localization are performed by identifying the peaks over all fast time bins and scan angles, yielding the estimated time delay index $\hat{n}_m$ and azimuth angle $\hat{\phi}_m$ for the $m^{\text{th}}$ target, as follows
\begin{equation}
(\hat{n}_m, \hat{\phi}_m) = \underset{n, \phi_i}{\arg\max} \; |\mathbf{\chi}[n, \phi_i]|^2.
\label{eq:peak_detect}
\end{equation}
For multiple targets, an ordered-statistic constant-false-alarm-rate (OS-CFAR) detector, with a false-alarm probability of $10^{-4}$, three training cells and one guard cell in range, and two training cells and one guard cell in azimuth, is applied to $|\chi[n,\phi_i]|^2$ over the complete range-azimuth map of $20$ range cells and $9$ azimuth cells. The range of the $m^{\text{th}}$ target is calculated using $r_m = \frac{c\, \hat{n}_m T_s}{2}$.

The convolution operations involved in time-domain matched filtering are computationally expensive for long sequences.
Hence, matched filtering is alternatively performed in the frequency domain. Here, the received signal $\mathbf{y}_d[n,\phi_i]$ is first Fourier transformed to the frequency domain,
\begin{equation}
\mathbf{Y}_d[k,\phi_i] = \mathcal{F}\{\mathbf{y}_d[n,\phi_i]\},
\label{eq:radar_rx_fft}
\end{equation}
followed by the Hadamard product of the received signal and the conjugate of the transmit Golay sequences in the frequency domain,
\begin{equation}
\begin{aligned}
\hat{\mathbf{\chi}}_d[k,\phi_i] = \mathbf{Y}_{d}[k,\phi_i] \odot \mathbf{G}_{d}^H[k].
\end{aligned}
\label{eq:xcorr_freq}
\end{equation}
The filtered outputs are then transformed back to the time domain and summed to obtain the combined range profile, as follows:
\begin{equation}
\mathbf{\chi}[n,\phi_i] = \mathcal{F}^{-1}\{\hat{\mathbf{\chi}}_{a}[k,\phi_i]\} + \mathcal{F}^{-1}\{\hat{\mathbf{\chi}}_{b}[k,\phi_i]\}.
\label{eq:radar_rx_ifft}
\end{equation}
The peak detection and range estimation are then performed as described in equation \eqref{eq:peak_detect}.

\subsection{Communication Signal Model and Processing}
Once the $m^{th}$ MU is localized in the range-azimuth space, the estimated azimuth angle, $\hat{\phi}_m$ is used to steer the transmit beam towards it for directional communication in Stage II. The communication is performed with one MU at a time in a TDM manner. The communication frame consists of a preamble, a header, and a data section. The preamble consists of a short training field and a long training field of $160$ samples each, where the short training field is used for the packet detection and the coarse frequency offset correction, and the long training field is used for the fine frequency offset correction and the channel estimation. The header, which carries the frame identifier, is mapped to BPSK symbols and transmitted in the first OFDM symbol, whereas the data is transmitted in the subsequent OFDM symbols. The communication supports BPSK, QPSK, 16-QAM, and 64-QAM with a fixed code rate of $1/2$, and only the mapping of the bits to the symbols of the Gray-coded constellation changes with the selected MCS. The bit-to-symbol mapping at the transmitter and the symbol-to-bit demapping at the receiver are performed in the PS. The data bits are digitally modulated to obtain symbols, $\mathbf{s}[q] \in \mathbb{C}^{1 \times Q}$, which are then assigned to $Q$ OFDM subcarriers. A $Q$-point IFFT is applied to obtain the time-domain OFDM signal. To mitigate inter-symbol interference, a cyclic prefix (CP) of length $N_{cp}$ is appended to each OFDM symbol by copying the last $N_{cp}$ samples of the time-domain OFDM signal, yielding the discrete-time sequence $\mathbf{x}_{cp}[n] \in \mathbb{C}^{1 \times (Q+N_{cp})}$, which is subsequently modulated onto the IF and converted to the analog domain via the DAC, where each sample is shaped by a root raised cosine pulse $\mathbf{w}(t)$ shifted by $nT_s$, yielding the continuous-time IF signal.
\begin{equation}
\mathbf{s}(t) = \sum_{n=-N_{\mathrm{cp}}}^{Q-1} \mathbf{x}_{cp}[n]\, \mathbf{w}(t - nT_s)\, e^{j2\pi f_{\mathrm{IF}} nT_s}
\label{eq:comm_if}
\end{equation}
In the testbed, $Q=64$ and $N_{cp}=16$, which results in $80$ samples per OFDM symbol. The continuous-time IF signal $\mathbf{s}(t)$ is then upconverted to the mmWave carrier frequency $f_c$. Finally, the signal is transmitted to the MU, through analog beamforming, along the pre-determined optimal beam $\hat{\phi}_m$.

At the MU-RX, the signal is captured across an $N_r$-element ULA. The down-converted, demodulated, and digitized received signal at the MU-RX is
\begin{equation}
    \mathbf{y}_c[n] = b_m\, G(\phi_m,\hat{\phi}_m)\, \mathbf{x}_c\left[n-\frac{n_m}{2}\right] + \mathbf{z}_c.
    \label{eq:comm_rx}
\end{equation}
where the transmitted signal is delayed by one-way propagation, and the one-way path loss is $b_m$, and $\mathbf{z}_c \in \mathbb{C}^{N_r \times 1}$ denotes the additive white Gaussian noise.
To maximize the SNR and suppress out-of-band noise, the received signal is first passed through a matched RRC filter $\mathbf{w}[n]$, corresponding to the transmit pulse shaping filter,
\begin{equation}
\mathbf{y}_{1}[n] = \mathbf{y}_c[n] \circledast \mathbf{w}[n].
\label{eq:comm_mf}
\end{equation}
Since the received signal's starting index is not known, the signal samples at the BS-RX are collected in a buffer. Then, autocorrelation is performed over a sliding window of length, $L_p$ as shown in
\begin{equation}
\mathcal{L}(\ell) = \left| \sum_{n=0}^{L_p-1} y_1[\ell+n]\, y_1^{*}[\ell+n-L_p] \right|^{2}, \quad \ell = L_p,\ldots,N-L_p.
\end{equation}
The packet start index $\hat{n}_s$ is identified as the index that
maximizes the normalized autocorrelation metric as
\begin{equation}
\hat{n}_s = \underset{\ell}{\arg\max} \; \dfrac{\mathcal{L}(\ell)}{\left( \sum_{n=0}^{L_p-1} \left| y_1[\ell+n-L_p] \right|^2 \right)^2}.
\label{eq:pd}
\end{equation}
A packet start is declared if the normalized metric at $\hat{n}_s$ exceeds
a predefined threshold $\gamma$, i.e.,
\begin{equation}
\dfrac{\mathcal{L}(\hat{n}_s)}{\left( \sum_{n=0}^{L_p-1} \left| y_1[\hat{n}_s+n-L_p] \right|^2 \right)^2} > \gamma.
\label{eq:threshold}
\end{equation}
Since the numerator $\mathcal{L}(\ell)$ and the square of the window energy in the denominator both have the units of volts to the fourth power, the normalized autocorrelation metric is dimensionless, and $\gamma$ is set to $0.8$. Then the coarse carrier frequency offset (CFO) $\Delta f$ is estimated from the phase derived of the autocorrelation between two consecutive short training field segments, each of length $L_s$, starting at $\hat{n}_s$, as follows:
\begin{equation}
\Delta\hat{f} = \frac{1}{2\pi L_s T_s} \arg \Biggl( \sum_{n=0}^{L_s-1} \mathbf{y}_{1}^*[\hat{n}_s+n] \cdot \mathbf{y}_{1}[\hat{n}_s+n+L_s] \Biggr)
\label{eq:cfo}
\end{equation}
The coarse CFO is then compensated by multiplying $\mathbf{y}_{1}[n]$ with a complex exponential, yielding the coarse corrected signal $\mathbf{y}_{2}[n] = \mathbf{y}_{1}[n]\, e^{-j2\pi \Delta\hat{f} nT_s}$. Subsequently, the residual fine CFO, $\Delta\hat{f}$, is estimated from the phase of the autocorrelation between two consecutive long training field segments, each of length $L_l$, as follows:
\begin{equation}
\delta \hat{f} = \frac{1}{2\pi L_l T_s} \arg \Biggl( \sum_{n=0}^{L_l-1} \mathbf{y}_2^*[\hat{n}_s+2L_s+n] \cdot \mathbf{y}_2[\hat{n}_s+2L_s+n+L_l] \Biggr)
\label{eq:ffo}
\end{equation}
The fine CFO correction is then applied to yield the fully synchronized signal $\mathbf{y}_3[n] = \mathbf{y}_2[n]\, e^{-j2\pi \delta\hat{f} nT_s}$. Following synchronization, the preamble comprising $2L_s$ short training field samples and $2L_l$ long training field samples, along with the cyclic prefix of length $N_{cp}$, are discarded from $\mathbf{y}_3[n]$. The $Q$ samples of the useful OFDM symbol are then extracted, yielding $\mathbf{y}_4[q] \in \mathbb{C}^{1 \times Q}$. A $Q$-point FFT is then applied to the $Q$ extracted samples of $\mathbf{y}_4[q]$, starting at index $\hat{n}_s + 2L_s + 2L_l + N_{cp}$ of $\mathbf{y}_3[n]$, to transform the signal to the frequency domain, yielding $\hat{\mathbf{y}}[k] \in \mathbb{C}^{1 \times Q}$, $\mathbf{Y}[k] = \mathcal{F}\{\mathbf{y}_4[n]\}$.

Further, channel estimation is performed in the frequency domain using known pilot symbols $\mathbf{S}_p[k] \in \mathbb{C}^{1 \times Q}$ transmitted on all $Q$ subcarriers. The estimated channel response,
\begin{equation}
\mathbf{H}[k] = \frac{\mathbf{Y}[k]}{\mathbf{S}_p[k]},
\label{eq:ch_est}
\end{equation}
captures the composite effect of the spatial channel gain $G(\phi_m, \hat{\phi}_m)$, the one-way path loss $b_m$, and the propagation delay across all $Q$ subcarriers.
Then, zero-forcing equalization is applied in the frequency domain,
by element-wise ratio of the received frequency domain signal $\mathbf{Y}[k]$ and the estimated channel response $\mathbf{H}[k]$, yielding
\begin{equation}
\mathbf{S}[k] = \frac{\mathbf{Y}[k]}{\mathbf{H}[k]}.
\label{eq:zf}
\end{equation}
The equalized symbols are then demapped based on the modulation constellation to recover the transmitted bits.

\section{Millimeter-Wave ISAC Hardware Testbed}
~\label{sec:testbed}
\begin{figure}[htbp]
    \centering
    \includegraphics[trim = 0 2.2cm 0 0, clip, width=0.5\textwidth]{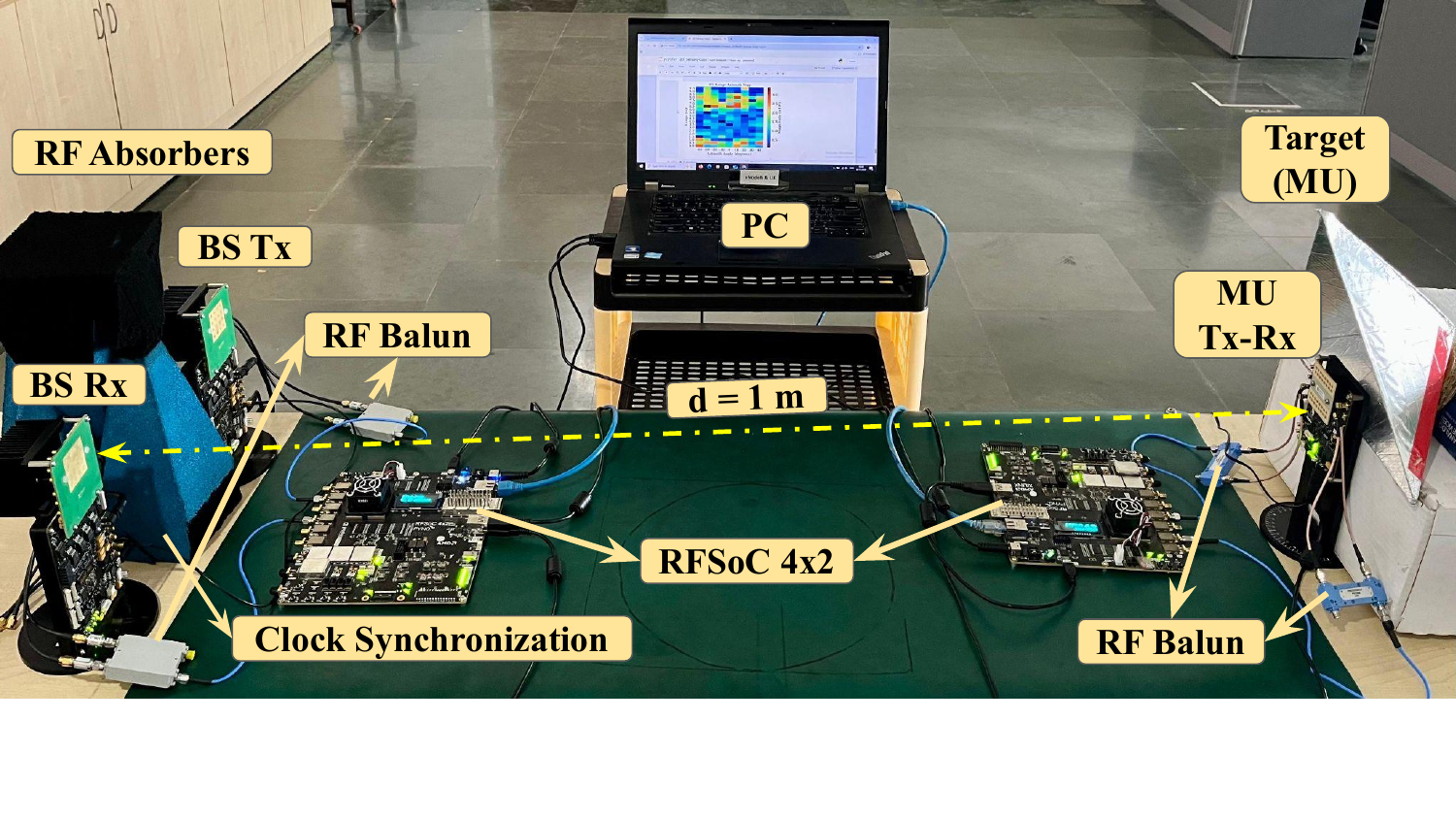}
    \caption{Proposed mmWave ISAC hardware testbed.}
    \label{fig:ISAC_Testbed}
\end{figure}

In this section, we present the specifications and design of the proposed mmWave ISAC testbed, along with the efficient mapping of the radar and communication signal processing algorithms onto the SoC through hardware-software co-design and word-length optimization. The proposed testbed, shown in Fig.~\ref{fig:ISAC_Testbed}, consists of two RFSoC 4x2 platforms, two Sivers EVK02004 AFEs, and one Sivers EVK02001 AFE. Each RFSoC integrates a multiprocessor system-on-chip (MPSoC) comprising a quad-core ARM Cortex-A53 as PS and a Zynq UltraScale+ FPGA as PL. For integration with the AFEs and over-the-air mmWave experiments, each RFSoC includes on-chip RF data converters (RFDCs) comprising digital up/down converters (DUCs/DDCs), analog-to-digital converters (ADCs), and DACs. Each RFSoC 4x2 board~\cite{RealDigitalRFSoC4x2} is built around the AMD Zynq UltraScale+ RFSoC XCZU48DR Gen 3 device~\cite{AMDDS889}, which consists of $14$-bit ADCs and DACs that operate at $4.9152$~GSPS in the testbed, while the PL is clocked at $307.2$~MHz. The waveform generation and the signal processing are performed in real time on the RFSoC, and no offline signal processing is performed on a PC.
The ISAC BS-TX transmits the ISAC waveform toward the MU, which also serves as the radar sensing target. The sensing target is realized by using a corner reflector placed at the MU location, and the BS-RX simultaneously processes the corresponding radar echo and the uplink communication signal from the MU under line-of-sight (LoS) conditions. At the BS, both radar and communication signals are processed by a common DFE and AFE, so a single front end supports both functions. The complete signal processing chain of the ISAC BS transceiver is shown in Fig.~\ref{fig:ISAC_BD}.

\subsection{Base Station Transmitter}
The proposed ISAC system operates in two stages: sensing and communication. In Stage 1, the BS-TX transmits the ISAC waveform, and the BS-RX uses radar sensing to estimate the MU's angular position. In Stage 2, the BS-TX transmits the information to be communicated via the ISAC waveform, using a directional communication link toward the MU.

\begin{figure*}[htbp]
    \centering
    \includegraphics[width=\textwidth]{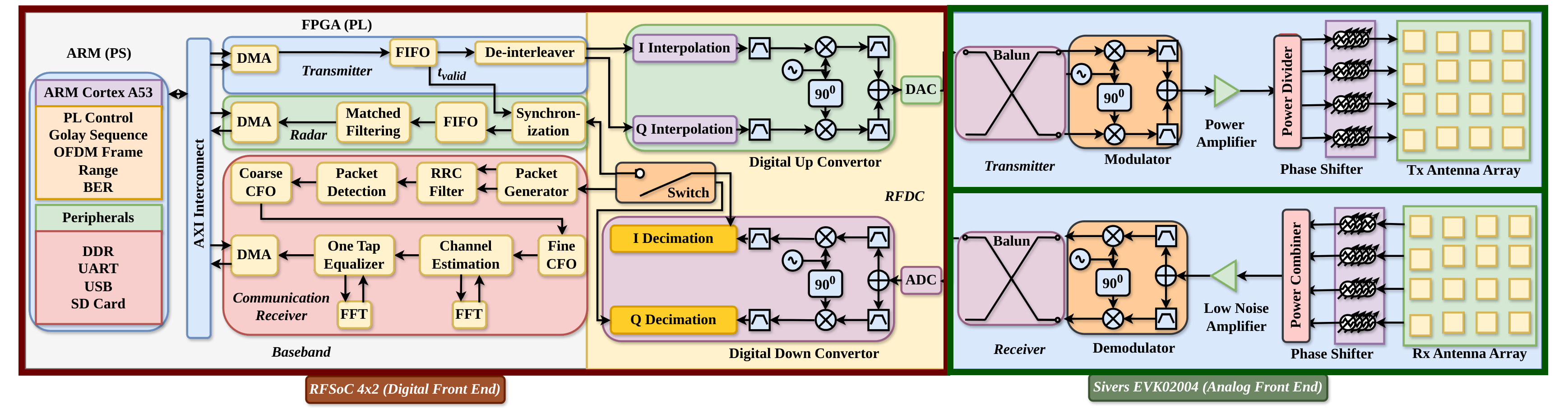}
    \caption{HSCD architecture of proposed BS ISAC at mmWave frequency with RFSoC 4x2 and Sivers EVK02004.}
    \label{fig:ISAC_BD}
\end{figure*}

The proposed architecture of the BS-TX and BS-RX is shown in Fig.~\ref{fig:ISAC_BD}. For the HSCD configuration, the BS-TX generates an IEEE 802.11ad frame in the PS, pulse-shapes it, and streams it to the RFDC over direct memory access (DMA) and the Advanced eXtensible Interface (AXI) protocol; a de-interleaver splits each complex sample into two real samples for the DAC, while the first-in-first-out  (FIFO) handles buffering across clock domains and supplies the valid signal used for radar receiver synchronization, as discussed in Section~\ref{sec:sync_cal}. The samples are scaled by $2^{14}$ to exploit the full DAC dynamic range, and since the PL clocks at 307.2 MHz, the built-in RFDC digital upconverter interpolates by 16 to the 4.9152 GSPS DAC rate and shifts the signal to an IF of 1.2 GHz using a numerically controlled oscillator; this IF is preferred over the 3.5 GHz upper limit of the EVK02004 upconverter since it provides 10 dB higher output power by avoiding the filter's band-edge attenuation. The DAC output drives the EVK02004 AFE through a Pasternack PE2CP1155 balun, which performs single-ended-to-differential conversion to reject common-mode signals at the AFE input. The AFE upconverts to 28.2 GHz using a mixer driven by a 27 GHz local oscillator, amplifies, and transmits over the air via analog beamforming from a $4 \times 4$ uniform rectangular array (URA) with half-wavelength spacing, $17^\circ$ beamwidth, and 25 dBi gain, yielding a 35 dBm effective isotropic radiated power  (EIRP) with electronic steering over $\pm40^\circ$ in azimuth and $\pm30^\circ$ in elevation in $10^\circ$ steps.

\begin{figure*}[t]
\centering
\includegraphics[width=\textwidth]{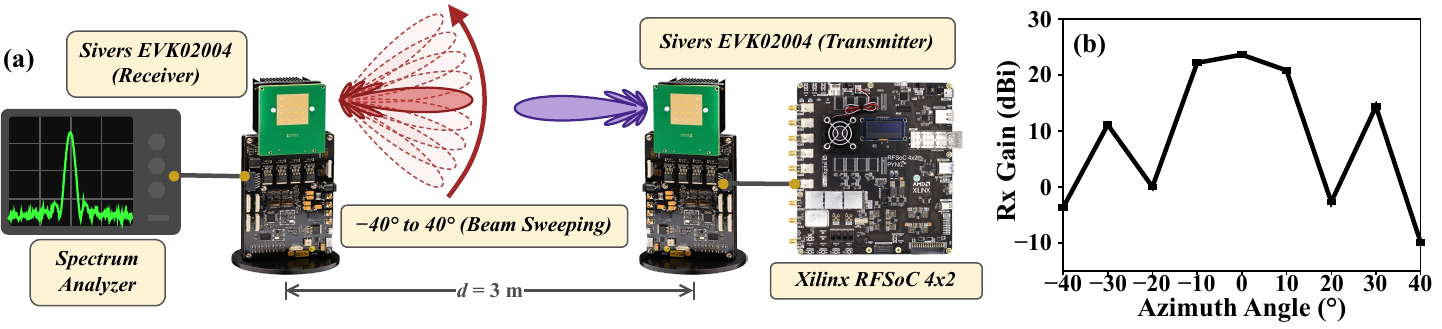}
\caption{(a) Experimental setup for the radiation-pattern measurement. (b) Measured antenna array gain of the $4\times4$ URA as a function of azimuth angle.}
\label{fig:radiation_pattern}
\end{figure*}

The gain of the URA is measured as a function of the steering angle. The antenna array gain patterns of the BS-TX and BS-RX URA are assumed to be identical as they are mounted on the Sivers platforms of identical make and model. Then, the gains are measured by fixing the distance between the transmitting and receiving antennas, the transmitted power, and the frequency of operation at 3 m, 10 dBm, and 28.2 GHz, respectively, across all the measurements, as shown in Fig.~\ref{fig:radiation_pattern}(a). In each measurement, the main beam from BS-TX is fixed at $0^{
\circ
}$, while the BS-RX beam is steered over $\pm40^\circ$ in steps of $10^\circ$, which is the minimum angular step supported by the hardware for analog beamforming. The received power is measured for each steering angle, from which the gain is estimated using the Friis's free space equation, as shown in Fig.~\ref{fig:radiation_pattern}(b).

\subsection{Base Station Receiver}
The BS-RX AFE uses a separate EVK02004, while its DFE shares the RFSoC 4x2 with the BS-TX; an external reference clock cable between the two EVK02004 modules keeps both the modules synchronized. The signal is captured by the $4 \times 4$ URA through analog beamforming, amplified by the LNA, demodulated to the 1.2 GHz IF, converted to single-ended by the PE2CP1155 balun, and digitized at 4.9152 GSPS. It is then demodulated to baseband by a numerically controlled oscillator, decimated by 16 to 307.2 MHz in the digital down-converter, and forwarded to the PL over the AXI stream interface. Full-duplex operation poses a further challenge, since direct coupling between the BS-TX and BS-RX antennas leaks the strong transmitted signal into the receive chain. We address this by placing pyramidal RF absorbers between the antennas, which yields a measured isolation gain of 27.8 dB.

\subsection{Radar Signal Processing}
Radar signal processing primarily comprises range estimation at a given azimuth via matched filtering. We consider two architectures for matched filtering: the first performs time-domain linear convolution, while the second performs frequency-domain matched filtering. We compare the two in terms of latency, computational complexity, resource utilization, and power consumption.

\subsubsection{Time Domain Linear Convolution Matched Filter}
\begin{figure}[htbp]
    \centering
    \includegraphics[width=0.5\textwidth]{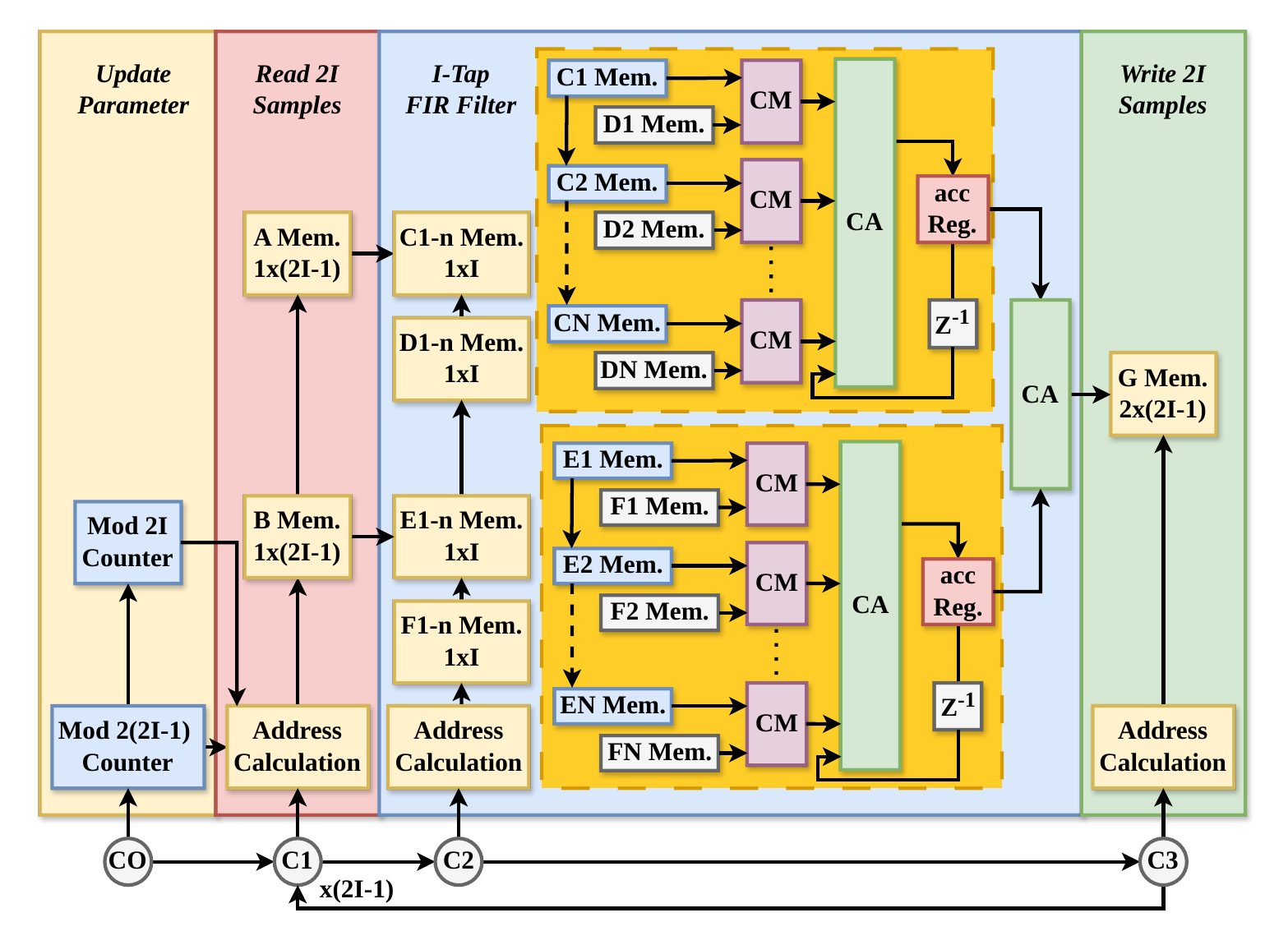}
    \caption{Hardware architecture of time-domain matched filtering.}
    \label{fig:MF_TD_IP}
\end{figure}
\begin{figure}[htbp]
    \centering\includegraphics[width=0.5\textwidth]{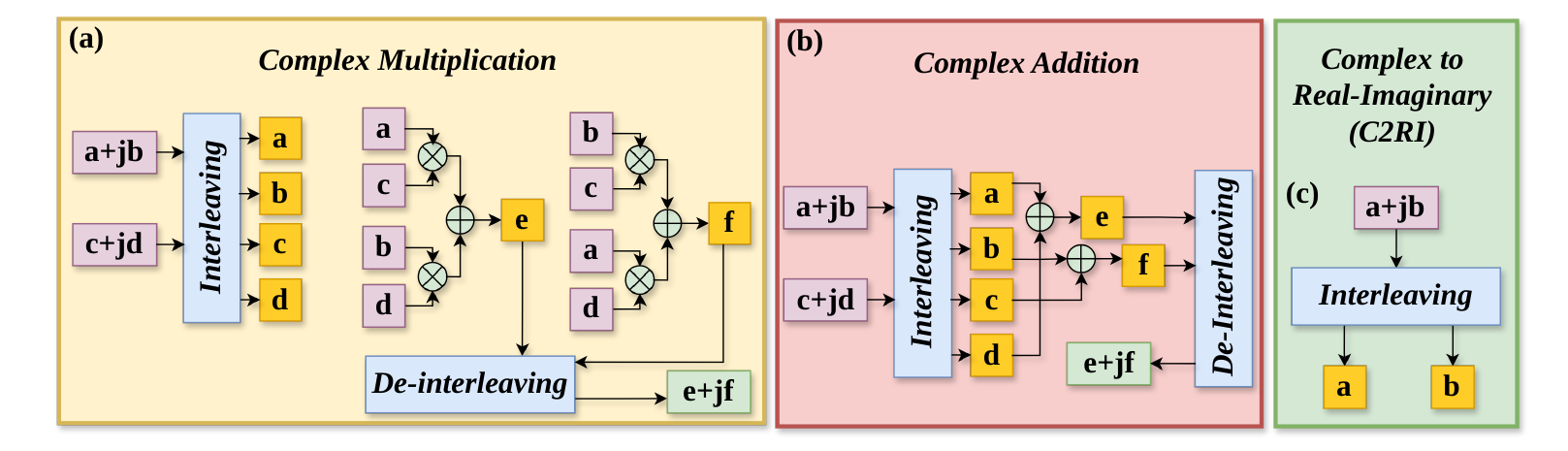}
    \caption{Hardware architecture of (a) complex multiplier, (b) complex adder, and (c) complex to real-imaginary.}
    \label{fig:Blocks_IP}
\end{figure}

As shown in equation \eqref{eq:xcorr}, matched filtering in the time domain is performed by correlating the received signal with each of the two complementary Golay transmit sequences. After the correlation is performed for both sequences, the outputs are coherently added to cancel the sidelobes, thereby improving the peak-to-sidelobe level ratio. As shown in Fig.~\ref{fig:MF_TD_IP}, the signal received from RFDC, $\mathbf{y}_a[n,\phi_i]$ and $\mathbf{y}_b[n,\phi_i]$ are stored in BRAMs $A$, and $B$, respectively. The corresponding transmit signal vectors, $\mathbf{g}^H_a[-n]$ and $\mathbf{g}^H_b[-n]$, are stored in BRAMs $D$ and $F$, respectively. The counters are initialized in the finite state machine (FSM) state $C0$. In state $C1$, the received signal vectors are read from BRAMs $A$ and $B$ and forwarded for the cross-correlation operation via intermediate BRAMs $C$ and $E$. In state $C2$, the $I$-element-wise complex multiply-and-accumulate (MAC) operations are performed between the received and transmit signal vectors. The cross-correlation outputs for both sequences are then summed using a complex adder. The accumulated values are stored in BRAM $G$ in state $C3$. The architectures for a complex multiplier, a complex adder, and an interleaver for complex-to-real mapping are shown in Fig.~\ref{fig:Blocks_IP}(a), (b), and (c), respectively.

\subsubsection{Frequency Domain Linear Convolution Matched Filter}

\begin{figure}[htbp]
    \centering\includegraphics[width=0.5\textwidth]{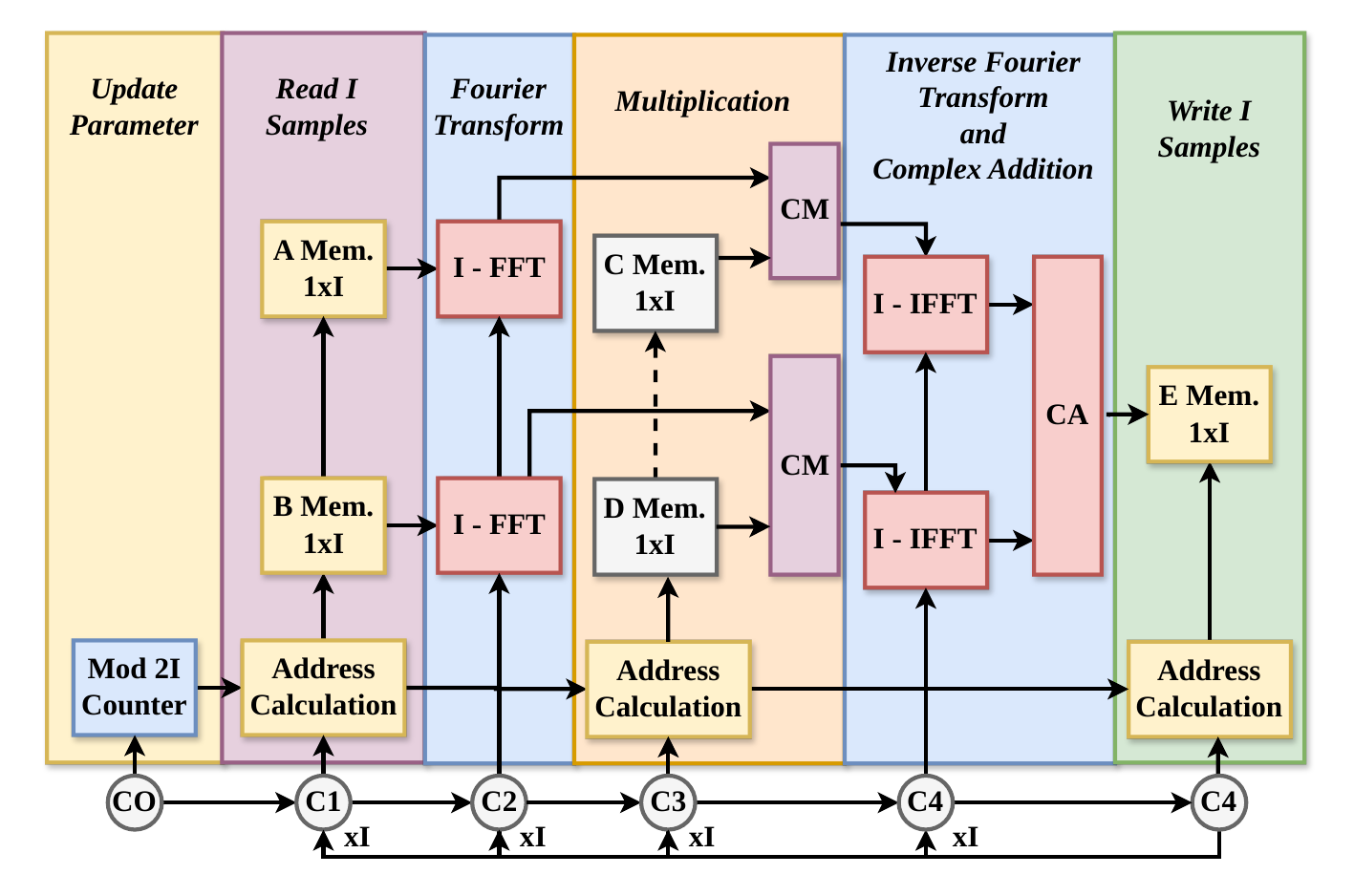}
    \caption{Hardware architecture of frequency-domain matched filtering. }
    \label{fig:MF_FD_IP}
\end{figure}

The computational complexity of the time-domain matched filtering architecture is $\mathcal{O}(N^2)$, which can be reduced to $\mathcal{O}(N\log N)$ by implementing it in the frequency domain. However, frequency-domain matched filtering requires three hardware units: an FFT, a multiplier, and an IFFT, which may result in additional communication overhead. The architecture is shown in Fig.~\ref{fig:MF_FD_IP}. The conjugate frequency-domain transmit signals $\mathbf{G}^H_a[k]$ and $\mathbf{G}^H_b[k]$ are pre-stored in BRAMs $C$ and $D$, respectively. In FSM state $C0$, the counter is initialized. In state $C1$, the received signals $\mathbf{y}_a[n,\phi_i]$ and $\mathbf{y}_b[n,\phi_i]$ are stored in BRAMs $A$ and $B$, respectively. As shown in equation \eqref{eq:radar_rx_fft}, an $I$-point FFT is performed in stage $C2$ to convert the time-domain signals into the frequency domain. In state $C3$, element-wise complex multiplication is performed between the frequency-domain received signals and the conjugate transmit signals, as presented in equation \eqref{eq:xcorr_freq}. In state $C4$, the multiplication outputs for both sequences are first converted back to the time domain by an $I$-point IFFT, equation \eqref{eq:radar_rx_ifft}, and the two time-domain outputs are then coherently added to cancel the sidelobes. The result is stored in BRAM $E$ in stage $C5$.

\subsubsection{Complexity Analysis}
Since frequency-domain matched filtering involves multiple hardware modules, we compare the resource utilization, execution time, and power consumption of single-precision floating-point (SPFL) and fixed-point word lengths on the PL, and against the PS. Here, a fixed-point word length of $<W,I>$ indicates total, integer, and fractional bits of $W$, $I$, and $(W-I)$, respectively. The lowest possible word length is identified such that there is no loss in functional accuracy of the hardware module relative to the floating-point output; hence, as shown in Column 2 of Table~\ref{tab:freq_comparison}, a different word length is selected for each module. To summarize, the fixed-point implementation results in a 59\% reduction in DSP units. Since the FPGA can use block RAM, flip-flops (FFs), and look-up tables (LUTs) for storage, the overall memory utilization of the fixed-point and SPFL architectures is nearly the same: the SPFL architecture favors FFs and LUTs, while the fixed-point architecture favors BRAMs. In terms of execution time, the fixed-point architecture is 17\% faster than the SPFL architecture on the PL, and it achieves a 20.5\% higher acceleration factor than SPFL relative to the PS implementation. Further, the fixed-point architecture consumes 5\% less power than SPFL.

Next, we compare the time-domain and frequency-domain matched filtering architectures in Table~\ref{tab:td_fd_comparison}. For the SPFL architectures, the frequency-domain architecture offers a substantial 53\% reduction in DSP units. Although it uses more BRAMs, its FF and LUT utilization is lower. It is also 34\% faster on the PS. When realized on the PL, the frequency-domain architecture is approximately 4\% faster than the time-domain architecture, with slightly higher power consumption. For the fixed-point architectures, the frequency-domain implementation consumes 4.79\% less power than the corresponding time-domain implementation and is 8.6\% faster, making the frequency-domain approach preferable for time-critical ISAC applications.

At the system level, the end-to-end RSP based on frequency-domain matched filtering takes 0.54~ms for each $\phi_i$, which corresponds to a 13.64\% reduction compared to the SPFL implementation and an acceleration factor of 9.95$\times$ over the PS implementation. Sweeping the full azimuth range from $-40^\circ$ to $+40^\circ$ at a $10^\circ$ step spans nine angles, giving a total processing time of approximately 5.22~ms per scan. In addition, an overhead is incurred for beam switching between consecutive angles, so that the total time for a complete scan exceeds the sum of the per-angle processing times. Including this beam-switching overhead, the measured time for a complete azimuth scan is approximately 0.58~s. For each beam, $0.58$~ms is spent on matched filtering, $4.4$~ms on DMA capture, and $59.46$~ms on switching the beam at both the BS-TX and the BS-RX, since the Sivers EVK02004 requires the register values of every beam to be separately written through the general-purpose input/output (GPIO) port of the RFSoC.

\begin{table*}[t]
\centering
\caption{Resource Utilization and Performance Comparison of Frequency Domain Receiver Architecture}
\label{tab:freq_comparison}
\renewcommand{\arraystretch}{1.2}
\setlength{\tabcolsep}{4pt}
\resizebox{\textwidth}{!}{%
\begin{tabular}{|c|c|c|c|c|c|c|c|c|c|c|c|c|c|c|c|c|}
\hline
\multirow{2}{*}{\textbf{Module}}
& \multirow{2}{*}{\makecell{\textbf{Word}\\\textbf{Length}}}
& \multicolumn{2}{c|}{\textbf{BRAM}}
& \multicolumn{2}{c|}{\textbf{FF}}
& \multicolumn{2}{c|}{\textbf{DSP}}
& \multicolumn{2}{c|}{\textbf{LUT}}
& \multirow{2}{*}{\textbf{PS (ms)}}
& \multicolumn{2}{c|}{\textbf{PL (ms)}}
& \multicolumn{2}{c|}{\textbf{Acceleration Factor}}
& \multicolumn{2}{c|}{\textbf{Power (W)}} \\
\cline{3-10} \cline{12-17}
&
& \textbf{SPFL} & \makecell{\textbf{Fixed Point}\\\textbf{(\% Gain)}}
& \textbf{SPFL} & \makecell{\textbf{Fixed Point}\\\textbf{(\% Gain)}}
& \textbf{SPFL} & \makecell{\textbf{Fixed Point}\\\textbf{(\% Gain)}}
& \textbf{SPFL} & \makecell{\textbf{Fixed Point}\\\textbf{(\% Gain)}}
&
& \textbf{SPFL} & \makecell{\textbf{Fixed Point}\\\textbf{(\% Gain)}}
& \textbf{SPFL} & \makecell{\textbf{Fixed Point}\\\textbf{(\% Gain)}}
& \textbf{SPFL} & \makecell{\textbf{Fixed Point}\\\textbf{(\% Gain)}} \\
\hline
FFT (1024 Point) & $<22,6>$
& 6 & \makecell{36\\($\uparrow$500.00\%)}
& 9936 & \makecell{3606\\(\textbf{$\downarrow$63.71\%})}
& 100 & \makecell{22\\(\textbf{$\downarrow$78.00\%})}
& 14005 & \makecell{4244\\(\textbf{$\downarrow$69.69\%})}
& 0.868
& 0.651 & \makecell{0.533\\(\textbf{$\downarrow$18.13\%})}
& 1.333 & \makecell{1.629\\(\textbf{$\uparrow$22.21\%})}
& 0.212 & \makecell{0.080\\(\textbf{$\downarrow$62.26\%})} \\
\hline
Multiplication & $<16,4>$
& 8 & \makecell{0\\(\textbf{$\downarrow$100.00\%})}
& 3169 & \makecell{1836\\(\textbf{$\downarrow$42.06\%})}
& 32 & \makecell{8\\(\textbf{$\downarrow$75.00\%})}
& 1963 & \makecell{2699\\($\uparrow$37.49\%)}
& 0.884
& 0.538 & \makecell{0.336\\(\textbf{$\downarrow$37.55\%})}
& 1.643 & \makecell{2.631\\(\textbf{$\uparrow$60.13\%})}
& 0.063 & \makecell{0.025\\(\textbf{$\downarrow$60.32\%})} \\
\hline
IFFT (1024 Point) & $<22,14>$
& 6 & \makecell{12\\($\uparrow$100.00\%)}
& 10458 & \makecell{8832\\(\textbf{$\downarrow$15.55\%})}
& 100 & \makecell{76\\(\textbf{$\downarrow$24.00\%})}
& 14209 & \makecell{15885\\($\uparrow$11.80\%)}
& 1.05
& 1.218 & \makecell{0.853\\(\textbf{$\downarrow$29.97\%})}
& 1.724 & \makecell{2.462\\(\textbf{$\uparrow$42.81\%})}
& 0.188 & \makecell{0.203\\($\uparrow$7.98\%)} \\
\hline
\textbf{Radar Module} & --
& \textbf{29} & \makecell{\textbf{57}\\($\uparrow$96.55\%)}
& \textbf{33647} & \makecell{\textbf{24357}\\(\textbf{$\downarrow$27.61\%})}
& \textbf{232} & \makecell{\textbf{95}\\(\textbf{$\downarrow$59.05\%})}
& \textbf{36036} & \makecell{\textbf{28685}\\(\textbf{$\downarrow$20.40\%})}
& \textbf{3.852}
& \textbf{0.654} & \makecell{\textbf{0.542}\\(\textbf{$\downarrow$17.13\%})}
& \textbf{5.89} & \makecell{\textbf{7.10}\\(\textbf{$\uparrow$20.54\%})}
& \textbf{2.923} & \makecell{\textbf{2.766}\\(\textbf{$\downarrow$5.37\%})} \\
\hline
\end{tabular}%
}
\end{table*}

\begin{table}[t]
\centering
\caption{Comparison of Time Domain vs Frequency Domain Receiver Architecture}
\label{tab:td_fd_comparison}
\renewcommand{\arraystretch}{1.2}
\setlength{\tabcolsep}{4pt}
\resizebox{\columnwidth}{!}{%

\begin{tabular}{|c|c|c|c|c|c|c|}
\hline

\multirow{2}{*}{\textbf{Parameter}}
& \multicolumn{3}{c|}{\textbf{SPFL}}
& \multicolumn{3}{c|}{\textbf{Fixed Point}}\\
\cline{2-7}

& \makecell[c]{\textbf{Time}\\\textbf{Domain}}
& \makecell[c]{\textbf{Frequency}\\\textbf{Domain}}
& \textbf{\% Gain}
& \makecell[c]{\textbf{Time}\\\textbf{Domain}}
& \makecell[c]{\textbf{Frequency}\\\textbf{Domain}}
& \textbf{\% Gain} \\
\hline

BRAM
& 13 & 29 & $\uparrow$123.08\%
& 10 & 57 & $\uparrow$470.00\% \\
\hline

FF
& 95316 & 33647 & \textbf{$\downarrow$64.70\%}
& 104239 & 24357 & \textbf{$\downarrow$76.63\%} \\
\hline

DSP
& 432 & 232 & \textbf{$\downarrow$53.70\%}
& 512 & 95 & \textbf{$\downarrow$81.45\%} \\
\hline

LUT
& 106151 & 36036 & \textbf{$\downarrow$66.05\%}
& 122316 & 28685 & \textbf{$\downarrow$76.55\%} \\
\hline

PS (ms)
& 5.86 & 3.852 & \textbf{$\downarrow$34.25\%}
& -- & -- & -- \\
\hline

PL (ms)
& 0.682 & 0.654 & \textbf{$\downarrow$4.11\%}
& 0.589 & 0.542 & \textbf{$\downarrow$7.98\%} \\
\hline

\makecell[c]{Accelration \\ Factor}
& 8.59 & 8.96 & \textbf{$\uparrow$4.3\%}
& 9.95 & 10.8 & \textbf{$\uparrow$8.6\%} \\
\hline

Power (W)
& 2.531 & 2.923 & $\uparrow$15.49\%
& 2.905 & 2.766 & \textbf{$\downarrow$4.79\%} \\
\hline

\end{tabular}%
}
\end{table}

\subsection{Range Estimation: Internal Delay Calibration}
\label{sec:sync_cal}
Accurate range estimation requires a precise estimate of the over-the-air propagation delay, but the signal also incurs an additional delay through the DFE and AFE that must be calibrated. One option is to connect the RFSoC TX and RX in loopback mode without the AFE; however, the measured delay fluctuated between 105 and 110 PL clock cycles across experiments due to a lack of TX-RX synchronization. At the 307.2 MHz PL clock, this corresponds to 0.34--0.35~$\mu$s, giving a range ambiguity of 2.5 m, five times the 0.5 m range resolution.
\begin{figure}[htbp]
    \centering
    \includegraphics[width=0.485\textwidth, trim=0cm 3cm 0cm 0cm, clip]{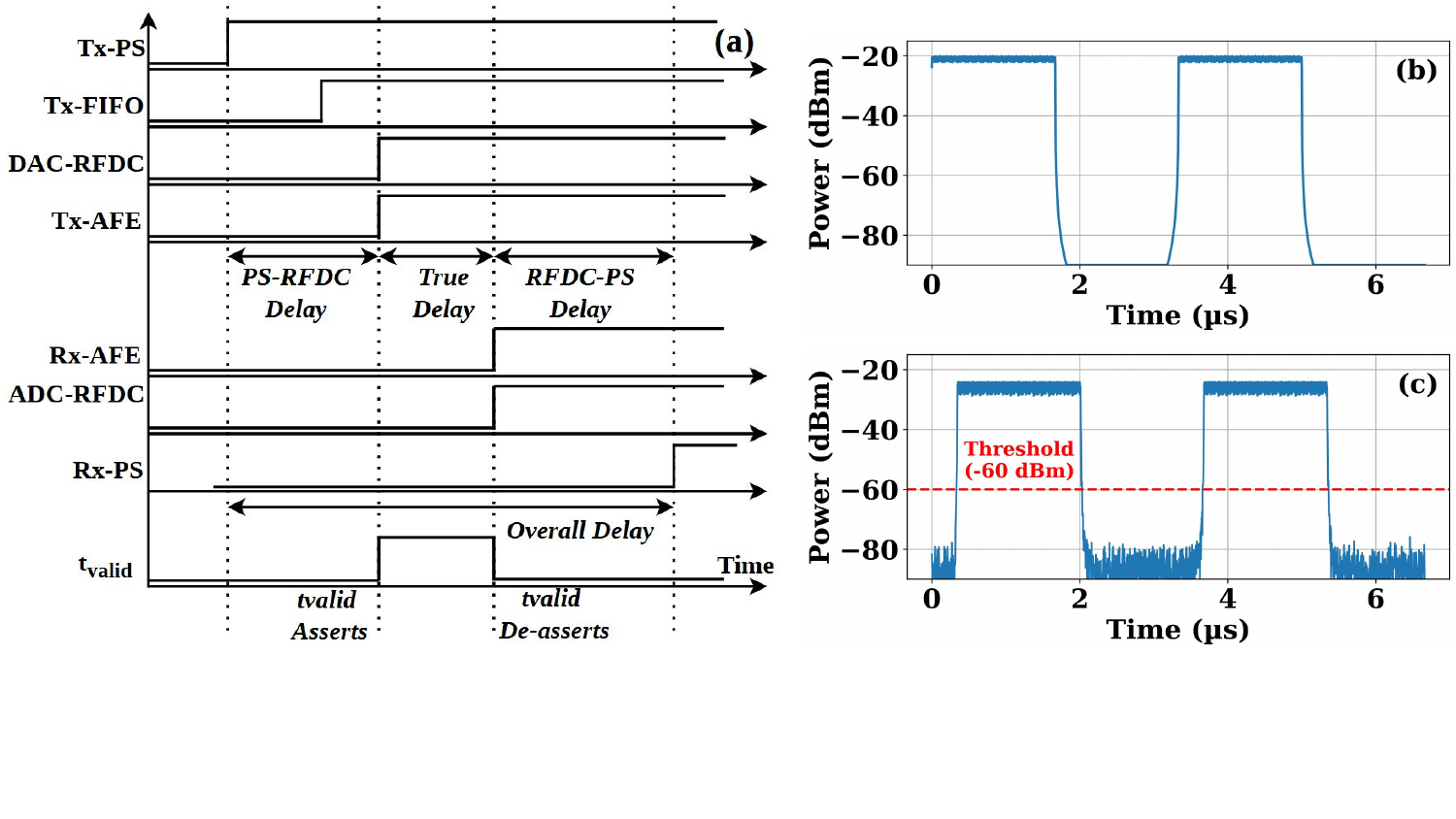}
    \caption{Synchronization for radar sensing: (a) Timing analysis, (b) transmit signal, (c) received signal.}
    \label{fig:sync}
\end{figure}
To address this issue, a synchronization module is included after the RFDC, as shown in Fig.~\ref{fig:ISAC_BD}, and a corresponding timing diagram explaining its operation is illustrated in Fig.~\ref{fig:sync}(a). At the BS-TX, the baseband samples are transferred from the PS to the RFDC via DMA through the TX-FIFO. Once the DMA begins streaming data into the TX-FIFO, the \texttt{tvalid} flag is asserted, marking the start of the timing reference. The signal then propagates through the DAC-RFDC and TX-AFE stages before being transmitted over the air. The cumulative delay before the \texttt{tvalid} assertion is denoted as the \textit{PS-RFDC Delay} in Fig.~\ref{fig:sync}(a), and comprises the DMA-to-RFDC delay ($A$) and the RFDC-to-AFE delay ($B$) on the transmit side.

At the BS-RX, the ADC is always ON and continuously digitizes the signal captured by the RX-AFE. The digitized samples first pass through the ADC-RFDC stage before reaching the PS. The delay introduced along this receive chain, from the physical arrival of the signal at the RX-AFE to its availability at the PS, is denoted as the \textit{RFDC-PS Delay}, and comprises the AFE-to-RFDC delay ($C$) and the RFDC-to-PS delay ($D$). The \textit{True Delay} is the actual round-trip over-the-air propagation delay between the BS and the target.

The \textit{Overall Delay} from the TX-PS to the RX-PS in Fig.~\ref{fig:sync}(a) consists of three parts, which are the \textit{PS-RFDC Delay} at the BS-TX, the delay through the RFDC, the AFEs, and the channel, and the \textit{RFDC-PS Delay} at the BS-RX. The \textit{PS-RFDC Delay} and the \textit{RFDC-PS Delay} include the PS and DMA transfers, which vary from one PRI to the next; hence, the \textit{Overall Delay} cannot be used for the range estimation. Instead, the delay is measured inside the PL as the duration of \texttt{tvalid}, which is de-asserted by the energy detector described below. This measured delay is $\tau_m = \tau_f + \tau_t$, where $\tau_t$ is the \textit{True Delay} and $\tau_f$ is the fixed latency of the RFDC and AFE stages between the two \texttt{tvalid} events. Since the PS and DMA delays lie outside this window, $\tau_f$ remains the same in every PRI and is measured only once using a calibration target at a known range $R_0$ as $\tau_f = \tau_{m,0} - 2R_0/c$, where $\tau_{m,0}$ is the measured delay for this target and $c$ is the speed of light. The range in every PRI is then obtained as $R = c(\tau_m - \tau_f)/2$.

To detect the transmitted signal at BS-RX, the synchronization module uses an energy-based threshold detector. As shown in Fig.~\ref{fig:sync}(b), the transmitted signal has a power level of approximately $-20$~dBm, while the received signal in Fig.~\ref{fig:sync}(c) additionally exhibits noise floor fluctuations and RF impairments. A detection threshold of $\zeta = -60$~dBm, set $10$~dB above the minimum signal of $-70$~dBm detectable by the ADC, is applied, and \texttt{tvalid} is de-asserted whenever the received power exceeds it for 10 consecutive samples, so that the duration for which \texttt{tvalid} remains asserted gives the measured delay $\tau_m$. Hence, the synchronization has two parts. First, the \texttt{tvalid} flag of the TX-FIFO is given to the RX to start the measurement of the delay, so that the TX and the RX share the same time reference. Second, the energy detector identifies the start of the received signal, which de-asserts \texttt{tvalid}, and only the samples from this instant are stored and passed on to the matched filtering. Realized on the PL close to the RFDC and repeated for every measurement, this approach absorbs small variations in internal delays and ensures that range estimation accuracy is bounded by the system's range resolution rather than by DFE timing jitter. Since the testbed uses a single DAC and a single ADC, multi-tile synchronization is not required.

\subsection{Communication Receiver: Baseband Signal Processing}
In this section, we present the detailed architecture of baseband signal processing for the BS-RX communication receiver, which is as shown in Fig.~\ref{fig:ISAC_BD}. It consists of: RRC filtering, packet detection, coarse and fine frequency synchronization, channel estimation, and one-tap equalization.

\subsubsection{RRC Filter}
\begin{figure}[htbp]
    \centering
    \includegraphics[width=0.5\textwidth]{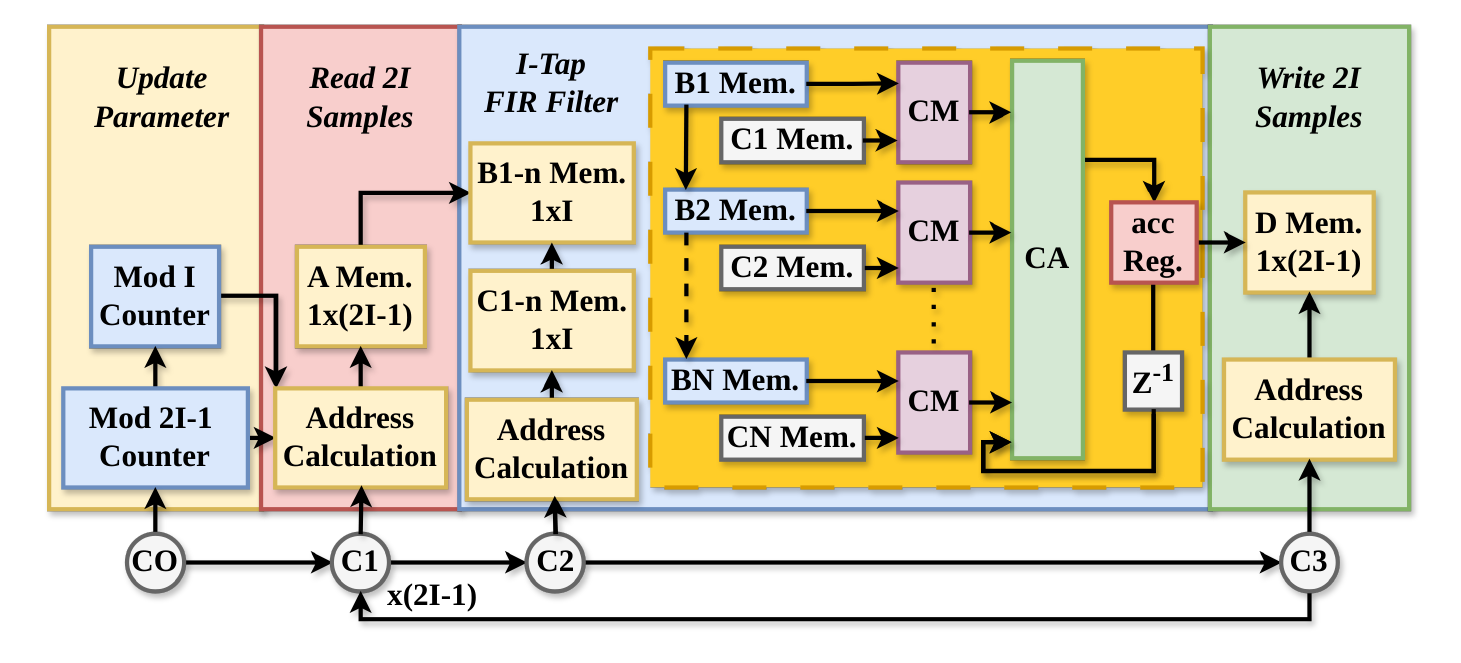}
    \caption{Hardware architecture of RRC filtering.}
    \label{fig:RRC_IP}
\end{figure}

RRC-based pulse-shape filtering is implemented as the first stage of the communication receiver, as shown in equation \eqref{eq:comm_mf}. The hardware architecture is realized as a finite impulse response (FIR) filter with a configurable roll-off factor. As shown in Fig.~\ref{fig:RRC_IP}, the counters are initialized in FSM state $C0$. In state $C1$, the received signal vectors $\mathbf{y}[n]$ are read from BRAM $A$, and the RRC filter coefficients $\mathbf{w}[n]$ are stored in BRAM $C$. In state $C2$, the received signal vectors and the RRC filter coefficients are read from BRAMs $B$ and $C$, respectively, and the cross-correlation between them is performed using the MAC unit. Finally, the output data $\mathbf{y}_{1}[n]$ is accumulated in a register and written to BRAM $D$ in state $C3$.

As shown in Table~\ref{tab:full_comparison}, the fixed-point realization of the RRC filter with a word length of $<29,4>$ offers 59\%, 61.16\%, and 6.7\% reductions in DSP, FF, and LUT usage, respectively, along with 26\% lower power consumption. Although it is 33\% slower than SPFL, it is 72$\times$ faster than the PS implementation, highlighting the importance of the PL. In the case of the RRC filter, the fixed-point and SPFL architectures are preferred for resource-constrained and time-constrained applications, respectively.

\subsubsection{Packet Detection}

\begin{figure}[htbp]
    \centering
    \includegraphics[width=0.5\textwidth]{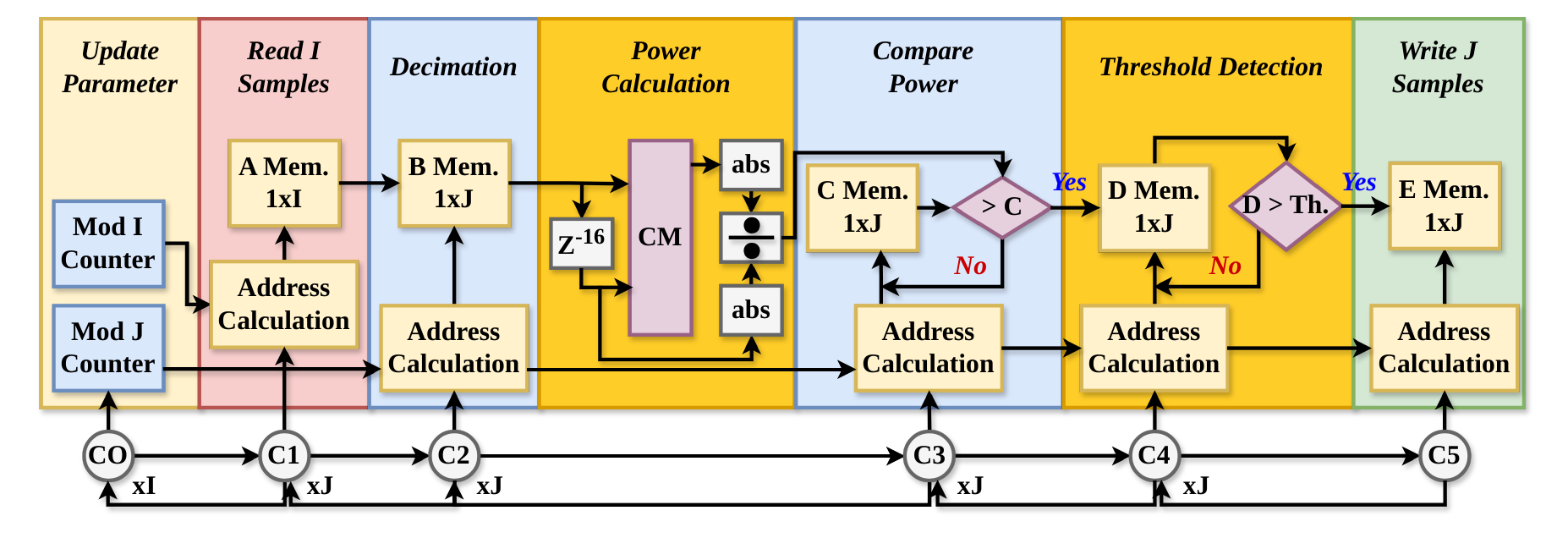}
    \caption{Hardware architecture of packet detection.}
    \label{fig:PD_IP}
\end{figure}

After RRC filtering, packet detection is performed as presented in equation \eqref{eq:pd}, and the corresponding architecture is shown in Fig.~\ref{fig:PD_IP}. After initializing the counter in FSM state $C0$, the RRC-filtered data vectors $\mathbf{y}_{1}[n]$ are stored in BRAM $A$ in state $C1$. In state $C2$, decimation is performed on the RRC-filtered data vectors, and the result is stored in BRAM $B$. The detection is based on an autocorrelation-based energy detection method, in which, for each sliding window index $\ell$, the autocorrelation between the current sample window $y_1[\ell+n]$ and its delayed counterpart $y_1[\ell+n-L_p]$ is computed over a window of length $L_p$. In state $C3$, the normalized autocorrelation values are compared with the preceding values stored in BRAM $C$, and the result is written to BRAM $D$. In state $C4$, the normalized autocorrelation ratio is compared against the threshold $\gamma$, and the detection flag is set if it exceeds $\gamma$, yielding the estimated packet start index $\hat{n}_s$. Finally, in state $C5$, the final output is written to BRAM $E$.

As shown in Table~\ref{tab:full_comparison}, the fixed-point realization of packet detection with a word length of $<24,6>$ offers 18\%, 39\%, 55\%, and 28\% reductions in BRAM, DSP, FF, and LUT, respectively, with 29\% lower power consumption. Although it is 26\% slower than SPFL, it is 21$\times$ faster than the PS implementation.

\begin{table*}[t]
\centering
\caption{Resource Utilization and Performance Comparison of Communication Receiver Architecture}
\label{tab:full_comparison}
\renewcommand{\arraystretch}{1.2}
\setlength{\tabcolsep}{4pt}
\resizebox{\textwidth}{!}{%
\begin{tabular}{|c|c|c|c|c|c|c|c|c|c|c|c|c|c|c|c|c|}
\hline
\multirow{2}{*}{\textbf{Module}}
& \multirow{2}{*}{\makecell{\textbf{Word}\\\textbf{Length}}}
& \multicolumn{2}{c|}{\textbf{BRAM}}
& \multicolumn{2}{c|}{\textbf{DSP}}
& \multicolumn{2}{c|}{\textbf{FF}}
& \multicolumn{2}{c|}{\textbf{LUT}}
& \multirow{2}{*}{\textbf{PS (ms)}}
& \multicolumn{2}{c|}{\textbf{PL (ms)}}
& \multicolumn{2}{c|}{\textbf{Acceleration Factor}}
& \multicolumn{2}{c|}{\textbf{Power (W)}} \\
\cline{3-10} \cline{12-17}
&
& \textbf{SPFL} & \makecell{\textbf{Fixed Point}\\\textbf{(\% Gain)}}
& \textbf{SPFL} & \makecell{\textbf{Fixed Point}\\\textbf{(\% Gain)}}
& \textbf{SPFL} & \makecell{\textbf{Fixed Point}\\\textbf{(\% Gain)}}
& \textbf{SPFL} & \makecell{\textbf{Fixed Point}\\\textbf{(\% Gain)}}
&
& \textbf{SPFL} & \makecell{\textbf{Fixed Point}\\\textbf{(\% Gain)}}
& \textbf{SPFL} & \makecell{\textbf{Fixed Point}\\\textbf{(\% Gain)}}
& \textbf{SPFL} & \makecell{\textbf{Fixed Point}\\\textbf{(\% Gain)}} \\
\hline

RRC Filter & $<29,4>$
& 0 & \makecell{0\\(--)}
& 410 & \makecell{168\\(\textbf{$\downarrow$59.02\%})}
& 35202 & \makecell{13674\\(\textbf{$\downarrow$61.16\%})}
& 32075 & \makecell{29915\\(\textbf{$\downarrow$6.73\%})}
& 77.43
& 0.81 & \makecell{1.07\\($\uparrow$33.75\%)}
& 91.13 & \makecell{72.37\\($\downarrow$20.60\%)}
& 0.771 & \makecell{0.57\\(\textbf{$\downarrow$26.07\%})} \\
\hline

Packet Detection & $<24,6>$
& 180.5 & \makecell{146.5\\(\textbf{$\downarrow$18.84\%})}
& 490 & \makecell{298\\(\textbf{$\downarrow$39.18\%})}
& 29558 & \makecell{13298\\(\textbf{$\downarrow$55.01\%})}
& 36626 & \makecell{26342\\(\textbf{$\downarrow$28.07\%})}
& 37.20
& 1.39 & \makecell{1.76\\($\uparrow$26.62\%)}
& 27.34 & \makecell{21.14\\($\downarrow$22.68\%)}
& 1.513 & \makecell{1.071\\(\textbf{$\downarrow$29.21\%})} \\
\hline

Coarse CFO & $<28,4>$
& 4 & \makecell{5\\($\uparrow$25.00\%)}
& 42 & \makecell{54\\($\uparrow$28.57\%)}
& 3860 & \makecell{1213\\(\textbf{$\downarrow$68.58\%})}
& 6139 & \makecell{4476\\(\textbf{$\downarrow$27.08\%})}
& 0.472
& 0.369 & \makecell{0.376\\($\uparrow$1.90\%)}
& 1.274 & \makecell{1.255\\($\downarrow$1.45\%)}
& 0.086 & \makecell{0.078\\(\textbf{$\downarrow$9.30\%})} \\
\hline

Fine CFO & $<31,6>$
& 6 & \makecell{3\\(\textbf{$\downarrow$50.00\%})}
& 42 & \makecell{43\\($\uparrow$2.38\%)}
& 3927 & \makecell{1353\\(\textbf{$\downarrow$65.55\%})}
& 6235 & \makecell{5230\\(\textbf{$\downarrow$16.12\%})}
& 0.472
& 0.369 & \makecell{0.376\\($\uparrow$1.90\%)}
& 1.274 & \makecell{1.255\\($\downarrow$1.45\%)}
& 0.091 & \makecell{0.096\\($\uparrow$5.49\%)} \\
\hline

FFT (64 Point) & $<16,6>$
& 30 & \makecell{14\\(\textbf{$\downarrow$53.33\%})}
& 48 & \makecell{40\\(\textbf{$\downarrow$16.67\%})}
& 37928 & \makecell{2198\\(\textbf{$\downarrow$94.20\%})}
& 21388 & \makecell{4176\\(\textbf{$\downarrow$80.48\%})}
& 0.16
& 0.37 & \makecell{0.368\\(\textbf{$\downarrow$0.54\%})}
& 0.438 & \makecell{0.435\\($\downarrow$0.69\%)}
& 0.263 & \makecell{0.098\\(\textbf{$\downarrow$62.74\%})} \\
\hline

Complex Division & $<26,5>$
& 6 & \makecell{6\\(--)}
& 20 & \makecell{4\\(\textbf{$\downarrow$80.00\%})}
& 9983 & \makecell{7654\\(\textbf{$\downarrow$23.32\%})}
& 3856 & \makecell{5655\\($\uparrow$46.65\%)}
& 0.43
& 0.008 & \makecell{0.003\\(\textbf{$\downarrow$87.50\%})}
& 53.75 & \makecell{143\\(\textbf{$\uparrow$166.04\%})}
& 0.093 & \makecell{0.12\\($\uparrow$29.03\%)} \\
\hline

\makecell[c]{\textit{$\hookrightarrow$ Channel Estimation}\\\textit{(FFT/2 + Division)}} & --
& \textit{21} & \makecell{\textit{13}\\(\textit{$\downarrow$38.10\%})}
& \textit{44} & \makecell{\textit{24}\\(\textit{$\downarrow$45.45\%})}
& \textit{28947} & \makecell{\textit{8753}\\(\textit{$\downarrow$69.76\%})}
& \textit{14550} & \makecell{\textit{7743}\\(\textit{$\downarrow$46.78\%})}
& \textit{0.51}
& \textit{0.193} & \makecell{\textit{0.187}\\(\textit{$\downarrow$3.11\%})}
& \textit{2.64} & \makecell{\textit{2.73}\\(\textit{$\uparrow$3.22\%})}
& \textit{0.2245} & \makecell{\textit{0.169}\\(\textit{$\downarrow$24.72\%})} \\
\hline

Complex Multiplication & --
& 0 & \makecell{--\\(--)}
& 26 & \makecell{--\\(--)}
& 3882 & \makecell{--\\(--)}
& 3842 & \makecell{--\\(--)}
& 0.22
& 0.005 & \makecell{--\\(--)}
& 44 & \makecell{--\\(--)}
& 0.101 & \makecell{--\\(--)} \\
\hline

\makecell[c]{\textit{$\hookrightarrow$ One-Tap Equalization}\\\textit{(FFT/2 + Multiplication)}} & --
& \textit{15} & \makecell{--\\(--)}
& \textit{50} & \makecell{--\\(--)}
& \textit{22846} & \makecell{--\\(--)}
& \textit{14536} & \makecell{--\\(--)}
& \textit{0.30}
& \textit{0.19} & \makecell{--\\(--)}
& \textit{1.58} & \makecell{--\\(--)}
& \textit{0.2325} & \makecell{--\\(--)} \\
\hline

\textbf{Communication Module} & --
& \textbf{272} & \makecell{\textbf{214}\\(\textbf{$\downarrow$21.32\%})}
& \textbf{1094} & \makecell{\textbf{633}\\(\textbf{$\downarrow$42.14\%})}
& \textbf{157518} & \makecell{\textbf{75852}\\(\textbf{$\downarrow$51.87\%})}
& \textbf{134883} & \makecell{\textbf{102651}\\(\textbf{$\downarrow$23.89\%})}
& \textbf{116.19}
& \textbf{2.29} & \makecell{\textbf{2.23}\\(\textbf{$\downarrow$2.62\%})}
& \textbf{51.09} & \makecell{\textbf{52.10}\\(\textbf{$\uparrow$1.98\%})}
& \textbf{7.44} & \makecell{\textbf{6.632}\\(\textbf{$\downarrow$10.86\%})} \\
\hline

\end{tabular}%
}
\end{table*}

\subsubsection{Coarse and Fine CFO Correction}

\begin{figure}[htbp]
    \centering
    \includegraphics[width=0.5\textwidth]{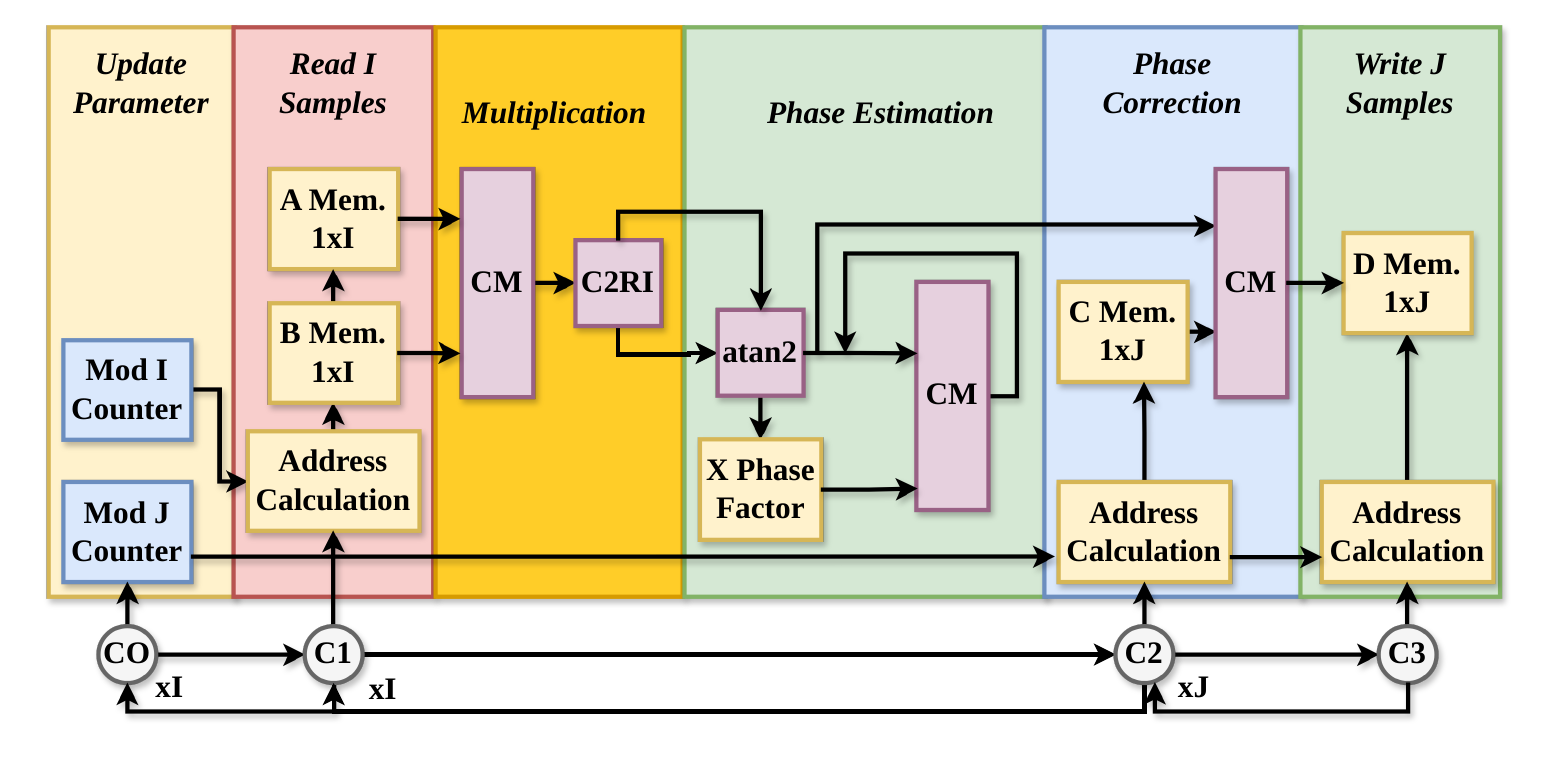}
    \caption{Hardware architecture of coarse and fine frequency offset.}
    \label{fig:CFO_FFO_IP}
\end{figure}

Once the packet start index $\hat{n}_s$ is identified, coarse and fine frequency offset corrections are performed. To estimate the coarse and fine frequency offsets, the short and long training fields in the preamble are used, respectively. Although both stages share an identical architecture, they are implemented as independent modules. In FSM state $C0$, the counters are initialized. In state $C1$, the first set of short training field samples of the received signal vector is stored in BRAM $A$, and the second set is stored in BRAM $B$. The complex conjugate multiplication is then performed, and the output is de-interleaved into the in-phase and quadrature components. The phase of the signal is then computed as $\tan^{-1}\!\left(\frac{Q}{I}\right)$, and the frequency offset is estimated by scaling the computed phase by the factor $\frac{1}{2\pi L_s T_s}$ for the coarse CFO and $\frac{1}{2\pi L_l T_s}$ for the fine CFO. The phase estimate is updated continuously for each sample. The received signal data vector is stored in BRAM $C$ in state $C2$, and phase correction is applied, as shown in equations \eqref{eq:cfo} and \eqref{eq:ffo}, by performing complex multiplication between the received signal data vectors and the estimated phase offset. In state $C3$, the output data is stored in BRAM $D$, as shown in Fig.~\ref{fig:CFO_FFO_IP}.

As shown in Table~\ref{tab:full_comparison}, the fixed-point realizations of the coarse and fine CFO modules do not offer any improvement over the SPFL architecture. Although keeping these two modules in the PS is a good alternative, the data movement between the PS and the PL leads to additional overhead, especially if the subsequent modules are in the PL.

\subsubsection{Channel Estimation}
\begin{figure}[htbp]
    \centering
    \includegraphics[width=0.5\textwidth]{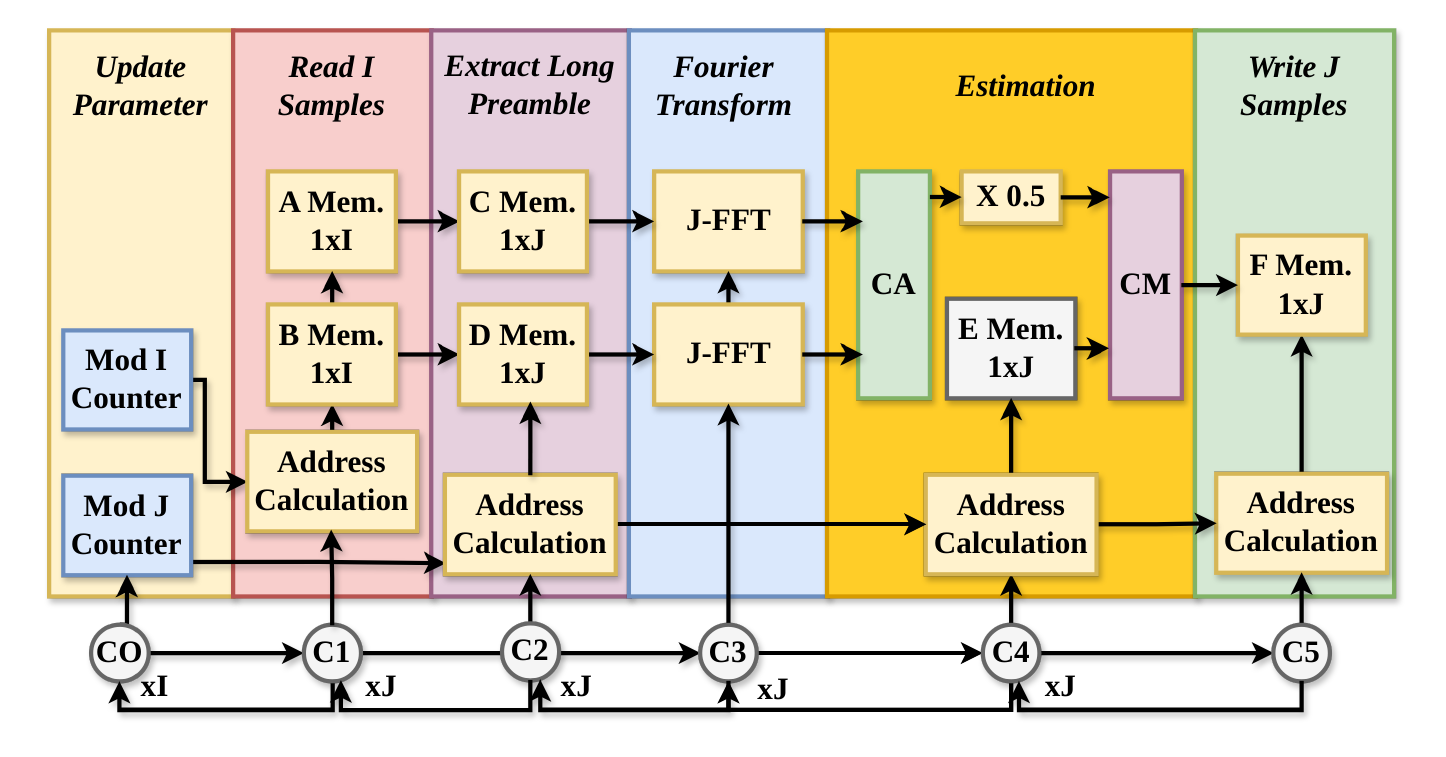}
    \caption{Hardware architecture of channel estimation in the frequency domain.}
    \label{fig:CH_EST_IP}
\end{figure}

Following synchronization, channel estimation is performed in the frequency domain using pilot symbols, as shown in equation \eqref{eq:ch_est}. In the FSM state $C0$, the counters are initialized. In state $C1$, the long training field samples at two different time offsets are stored in BRAMs $A$ and $B$. In state $C2$, the long training field is extracted and stored in BRAMs $C$ and $D$. A $J$-point FFT is then performed on the long training field samples in state $C3$. The known pilot samples $\mathbf{s}_p[k]$ are pre-stored in BRAM $E$. Least-squares channel estimation is then performed by multiplying the frequency-domain long training field samples $\hat{\mathbf{y}}[k]$ by the complex conjugate of the known pilot symbols $\mathbf{s}_p[k]$ using a complex multiplier block. In state $C4$, the two resulting estimates are combined via a complex adder block with a scaling factor of $0.5$ to obtain the averaged channel estimate $\hat{\mathbf{h}}[k]$. The estimate is then stored in BRAM $F$ in state $C5$, as shown in Fig.~\ref{fig:CH_EST_IP}.

The channel estimation block, comprising the $J$-point FFT and the complex multiplier, is summarized in Table~\ref{tab:full_comparison}. Compared with SPFL, the fixed-point implementation offers reductions of 38.10\%, 45.45\%, 69.76\%, and 46.78\% in BRAM, DSP, FF, and LUT utilization, respectively, along with 24.7\% lower power consumption, and is 3.22\% faster.

Since the FFT is used in matched filtering in the RSP, channel estimation, and equalization, we compare the performance of the built-in FFT from AMD-Xilinx with the proposed implementation for the communication word length of $\langle 16,6 \rangle$. As shown in Table~\ref{tab:FFT_WLO_HLS}, the proposed architecture offers significant reductions in FF and LUT usage, as well as in execution time and power consumption. Even though there is a marginal increase in BRAM and DSP usage, the proposed fixed-point architecture is a better alternative. Thus, although the built-in FFT provided by AMD-Xilinx is efficient for SPFL and widely used, custom architectures are preferred for arbitrary word lengths, highlighting the importance of word-length analysis.

\begin{table}[ht]
\centering
\caption{Resource Utilization and Performance Comparison of FFT Implementations}
\begin{tabular}{|c|c|c|c|}
\hline
\textbf{Parameter} & $\textbf{FFT}\langle 16,6 \rangle$ & \textbf{FFT (HLS)} & \textbf{\% Gain} \\
\hline
BRAM & 7 & 3 & \textbf{$\downarrow$ 57.14\%} \\
\hline
FF & 1099 & 28253 & $\uparrow 2470.79\%$ \\
\hline
DSP & 20 & 12 & \textbf{$\downarrow$ 40.00\%} \\
\hline
LUT & 2088 & 10921 & $\uparrow 422.84\%$ \\
\hline
PS Time (ms) & 0.16 & 0.16 & --- \\
\hline
PL Time (ms) & 0.368 & 0.371 & $\uparrow 0.82\%$ \\
\hline
\makecell[c]{Accelration \\ Factor}& 0.434 & 0.431 & $\downarrow 0.69\%$ \\
\hline
Power & 0.041 & 0.17 & $\uparrow 314.63\%$ \\
\hline
\end{tabular}
\label{tab:FFT_WLO_HLS}
\end{table}

\subsubsection{One Tap Equalizer}
At the final stage of the communication receiver chain on the PL, one-tap zero-forcing equalization is implemented to recover the transmitted data symbols from the frequency-domain received signal, as shown in equation \eqref{eq:zf}. The counter is initialized in FSM state $C0$. The received signal at two different time offsets is stored in BRAMs $A$ and $B$ in state $C1$. In state $C2$, the CP is removed to recover the data symbols, which are stored in BRAMs $C$ and $D$. A $J$-point FFT is then performed on the received data symbols to convert them to the frequency domain in state $C3$. The channel estimates $\hat{\mathbf{h}}[k]$ are stored in BRAMs $E$ and $F$. In state $C4$, zero-forcing equalization is performed by first computing the reciprocal of the estimated channel response $\hat{\mathbf{h}}[k]$ and then multiplying it with the frequency-domain received samples $\hat{\mathbf{y}}[k]$ using a complex multiplication block, yielding the equalized symbols $\hat{\mathbf{s}}[k]$. Finally, in state $C5$, the equalized data payloads are appended with their corresponding indices and written sequentially to BRAM $G$.

As shown in Table~\ref{tab:OTE_SPFL_WL}, the fixed-point architecture does not offer any substantial gain over the SPFL architecture other than in DSP usage. Hence, the SPFL architecture is preferred, and the PL is preferred over the PS owing to an acceleration factor of 44$\times$.

\begin{figure}[htbp]
    \centering
    \includegraphics[width=0.5\textwidth]{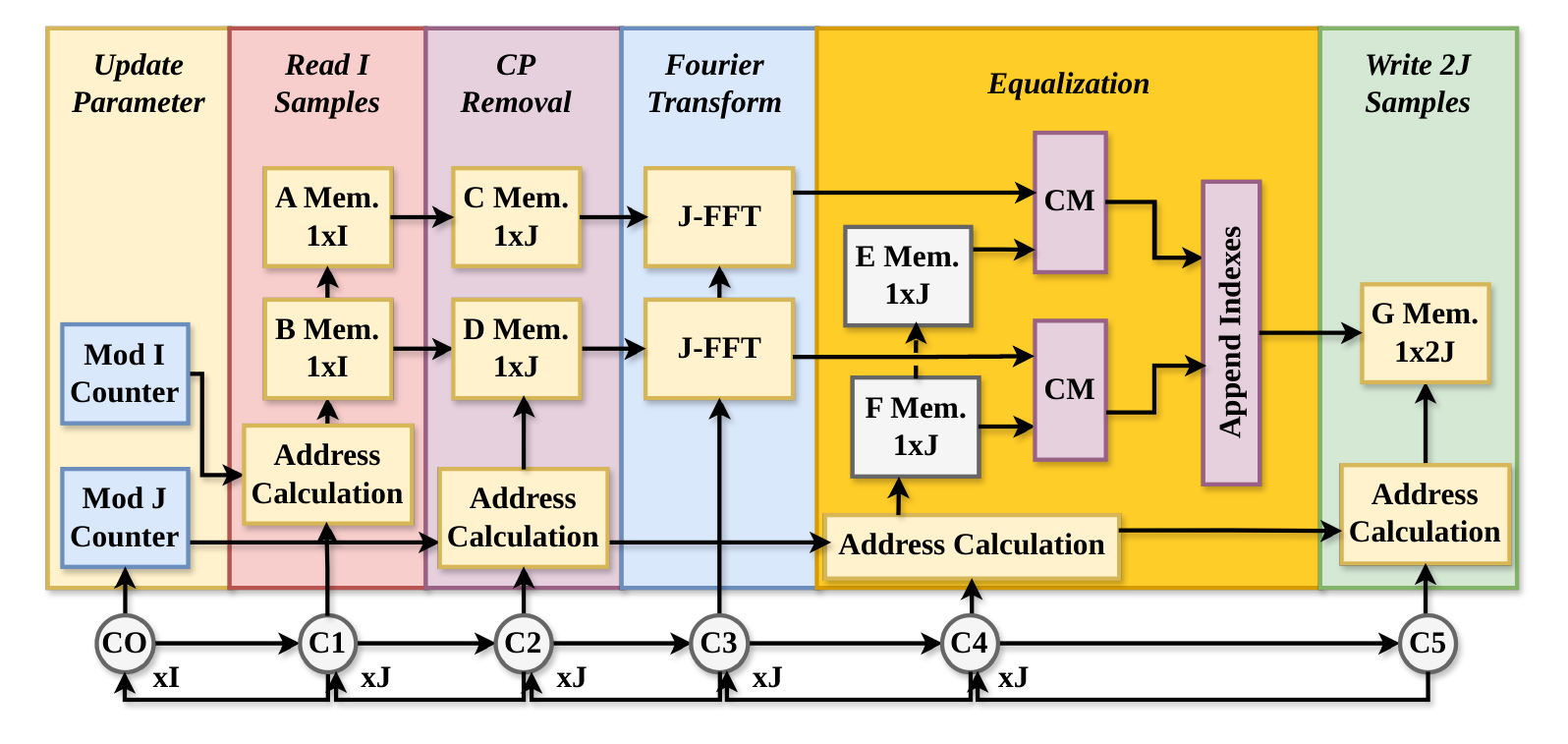}
    \caption{Architecture of one tap equalization in the frequency domain.}
    \label{fig:OTE_IP}
\end{figure}

\begin{table}[ht]
\centering
\caption{Comparison Between SPFL and Fixed Point Implementation of One Tap Equalizer}
\begin{tabular}{|c|c|c|c|}
\hline
\textbf{Parameter} & \textbf{SPFL} & $\textbf{Fixed Point }\langle 29,14 \rangle$ & \textbf{\% Gain} \\
\hline
BRAM & 6 & 6 & --- \\
\hline
FF & 4476 & 4346 & \textbf{$\downarrow$ 2.90\%} \\
\hline
DSP & 42 & 20 & \textbf{$\downarrow$ 52.38\%} \\
\hline
LUT & 5542 & 21898 & $\uparrow 295.13\%$ \\
\hline
PS Time (ms) & 0.22 & 0.22 & --- \\
\hline
PL Time (ms) & 0.005 & 0.004 & \textbf{$\downarrow$ 20.00\%} \\
\hline
\makecell[c]{Acceleration \\ Factor} & 44 & 55 & \textbf{$\uparrow$ 25.00\%} \\
\hline
Power & 0.145 & 0.551 & $\uparrow 280.00\%$ \\
\hline
\end{tabular}
\label{tab:OTE_SPFL_WL}
\end{table}

\subsubsection{Complexity Comparison}
The last row of Table~\ref{tab:full_comparison} summarizes the resource utilization, processing latency, acceleration factor, and power consumption for both the SPFL and fixed-point implementations of the end-to-end communication receiver. The fixed-point implementation consistently reduces the hardware resource overhead, achieving total reductions of 21.32\%, 42.14\%, 51.87\%, and 23.89\% in BRAM, DSP, FF, and LUT utilization, respectively, compared with the SPFL implementation. In terms of performance, the fixed-point design achieves a total acceleration factor of 52.10$\times$ over the PS software execution time, at a PL execution time of 2.23~ms and a total board power consumption of 6.632~W. These correspond to improvements of 1.98\%, 2.62\%, and 10.86\% over the SPFL baseline in acceleration factor, execution time, and power consumption, respectively. These results validate the fixed-point HSCD architecture as a resource-efficient and low-power solution for real-time communication-receiver processing in mmWave ISAC systems.

\subsection{Mobile User Transceiver}
The MU serves as a communication-only node and does not support radar signal processing. The MU-TX and MU-RX share the same Xilinx RFSoC 4x2 platform, which also hosts the DFE, with a communication hardware implementation identical to that of the ISAC BS transceiver. The primary hardware difference between the MU and the BS lies in the AFE: the MU uses the Sivers EVK02001, which is equipped with an $8 \times 2$ URA providing an antenna gain of 25 dBi. It supports electronic beam steering over $\pm45^\circ$ in azimuth with a step size of $4.5^\circ$, offering finer angular resolution than that of the BS for communication beam alignment. In the experiments, the MU sweeps $9$ beams over $\pm36^\circ$ in steps of $9^\circ$. We have omitted the architecture details due to space constraints.

\section{Results and Analysis}
\label{sec:results}
The proposed ISAC testbed is evaluated across three dimensions: radar sensing performance, measured through range and angle estimation accuracy in single- and multiple-target scenarios; communication link quality, in terms of EVM and BER across different modulation orders, bandwidths, and distances; and the interdependence between radar and communication. All experiments are conducted over-the-air in real indoor radio environments, inside laboratories and corridors, rich in multipath scattering.

Before the testbed is used for experimental validation of radar and communication performance, it must be calibrated. The noise floor at the ADC input of the BS-RX is measured with a spectrum analyzer as $-90$~dBm when the BS-RX is off and $-80$~dBm when it is on. Over the $307.2$~MHz processing bandwidth, this gives a noise figure of $9.1$~dB for the BS-RX chain. The minimum signal detectable by the ADC is $-70$~dBm, and the BS-RX enters saturation when the signal power exceeds $-20$~dBm. Prior to each measurement, the variable-gain amplifier is configured to maintain the received signal level within the system's dynamic range throughout the unambiguous sensing range of 256 m.
\begin{figure}[htbp]
    \centering
    \includegraphics[width=0.45\textwidth, trim= 0cm 0cm 4.5cm 0cm, clip]{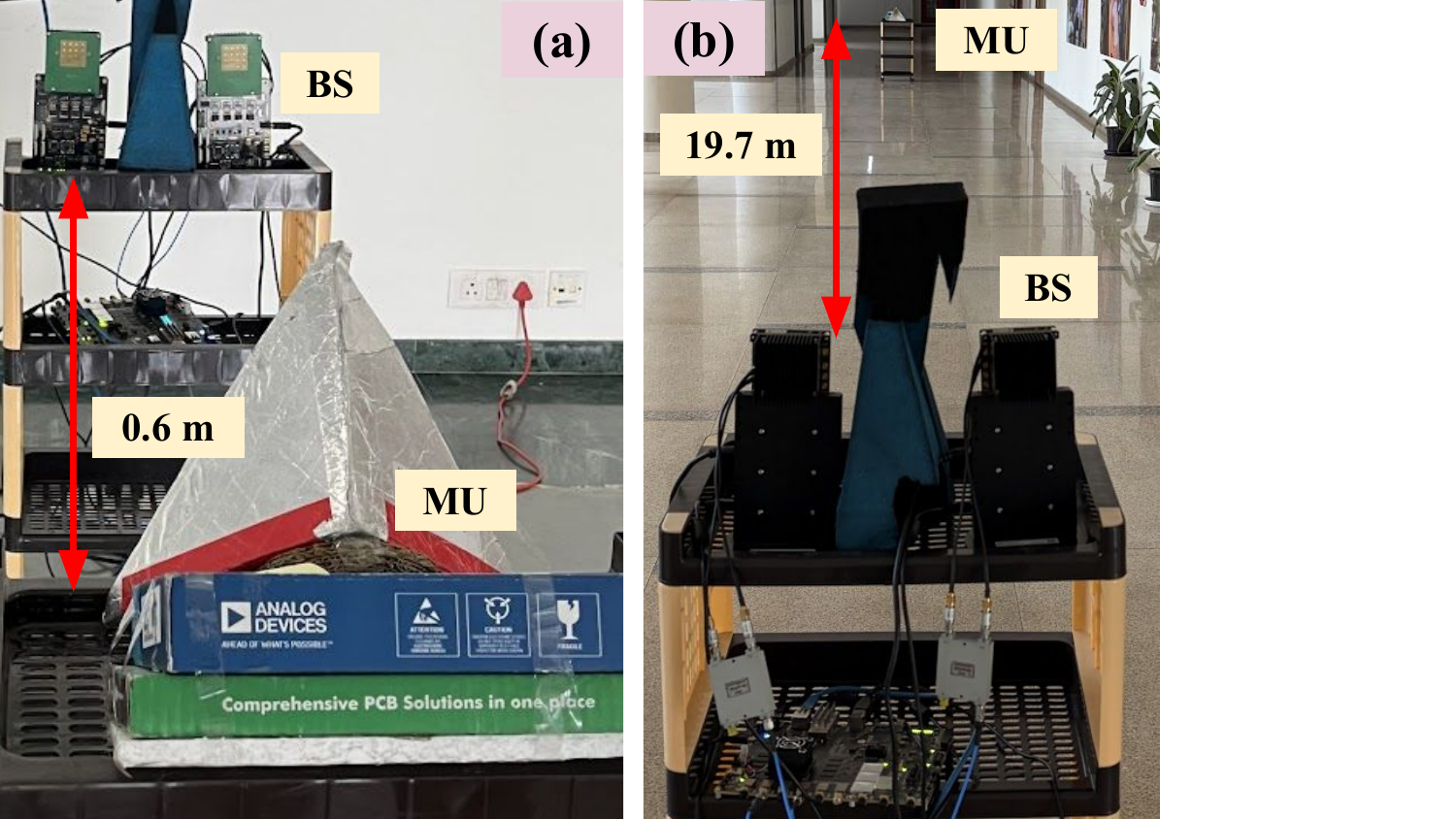}
    \caption{Experimental setup for radar range estimation: (a) Range = 0.6 m, (b) Range = 19.7 m}
    \label{fig:radar_range_setup}
\end{figure}
\subsection{Radar Performance Characterization}
In the first set of experiments, the BS-TX and BS-RX are configured for monostatic radar operation. The radar's performance is evaluated for range and angle estimation in single- and multiple-target scenarios.
\begin{figure}[htbp]
    \centering
    \includegraphics[width=0.5\textwidth]{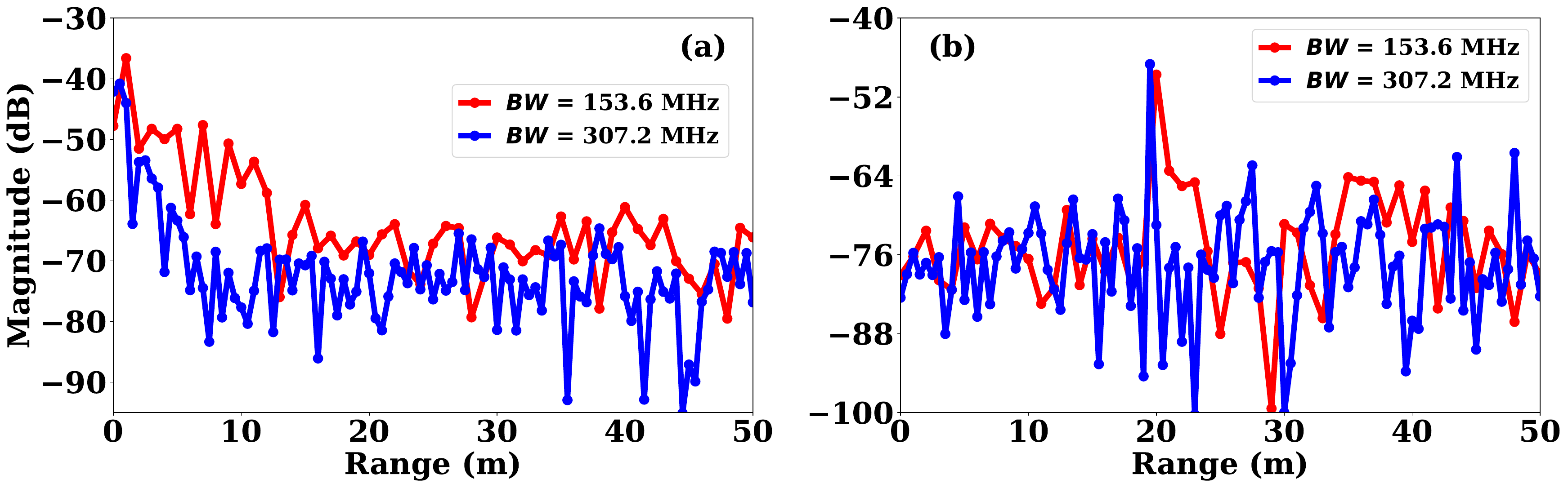}
    \caption{Range estimation using matched filter for target at range: (a) 0.6 m, and (b) 19.7 m.}
    \label{fig:MF_Output}
\end{figure}
The range estimation is evaluated by placing a single trihedral corner reflector with triangular faces of inner edge length $19$~cm and a radar cross-section (RCS) of $16.8$~dBsm directly in front of the radar at 0$^\circ$ azimuth and varying the range between 0.6 m and 19.7 m as shown in Fig.~\ref{fig:radar_range_setup}(a) and (b), respectively. The corresponding high range resolution profiles are shown in Fig.~\ref{fig:MF_Output}(a) and (b). The target ranges are estimated at the peaks of the profiles at 1 m and 20 m, for each case. This corresponds to absolute errors of 0.4 m and 0.3 m, for a 153.6 MHz bandwidth. The experiments are then repeated with a bandwidth of 307.2 MHz, and the target ranges are estimated at 0.5 m and 19.5 m, corresponding to absolute errors of 0.1 m and 0.2 m, in Figs.~\ref{fig:MF_Output}(a) and (b), respectively.
\begin{figure}[htbp]
    \centering
    \includegraphics[width=0.5\textwidth]{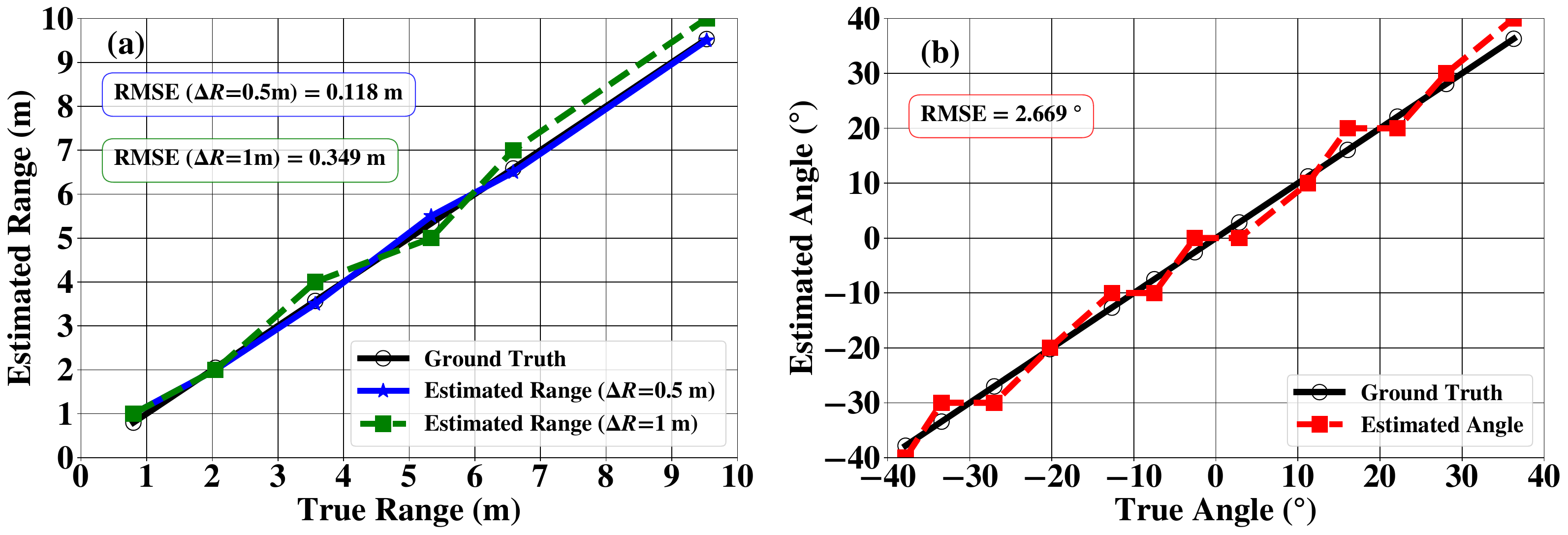}
    \caption{RMSE comparison for (a) range, and (b) azimuth.}
    \label{fig:RA_RMSE}
\end{figure}
To quantify estimation accuracy across multiple target positions, the experiment was then extended to 20 target positions spanning 0.5 m to 10 m in steps of 0.5 m, for both bandwidth configurations. For each position, the RMSE is computed between the ground truth target's range, $x_r$, and the estimated range, $\hat{x}_r$,
\begin{equation}
\text{RMSE} = \sqrt{\frac{1}{R} \sum_{r=1}^{R} \left( \hat{x}_r - x_r \right)^2}.
\label{eq:rmse}
\end{equation}
The range RMSE is 0.11 m for 307.2 MHz and 0.349 m for 153.6 MHz, as shown in Fig.~\ref{fig:RA_RMSE}(a), which is below $\frac{c}{2BW}$, where c is the speed of light, and BW is the bandwidth of the radar signal. The range-estimation accuracy also reflects the accuracy of the proposed synchronization module to estimate DFE-induced delay. In the FR2-1 range of $24.25$-$52.6$~GHz, the maximum channel bandwidth is $400$~MHz~\cite{3gpp.38.101-2}, which is also the limit of the Sivers AFE. A processing bandwidth of $307.2$~MHz is used to keep a margin for the leakage outside the band, which allows a range resolution of 0.488 m. The range accuracy is limited by the bandwidth of the AFE and not by the signal processing on the RFSoC.
\begin{figure}[htbp]
    \centering
    \includegraphics[width=0.5\textwidth]{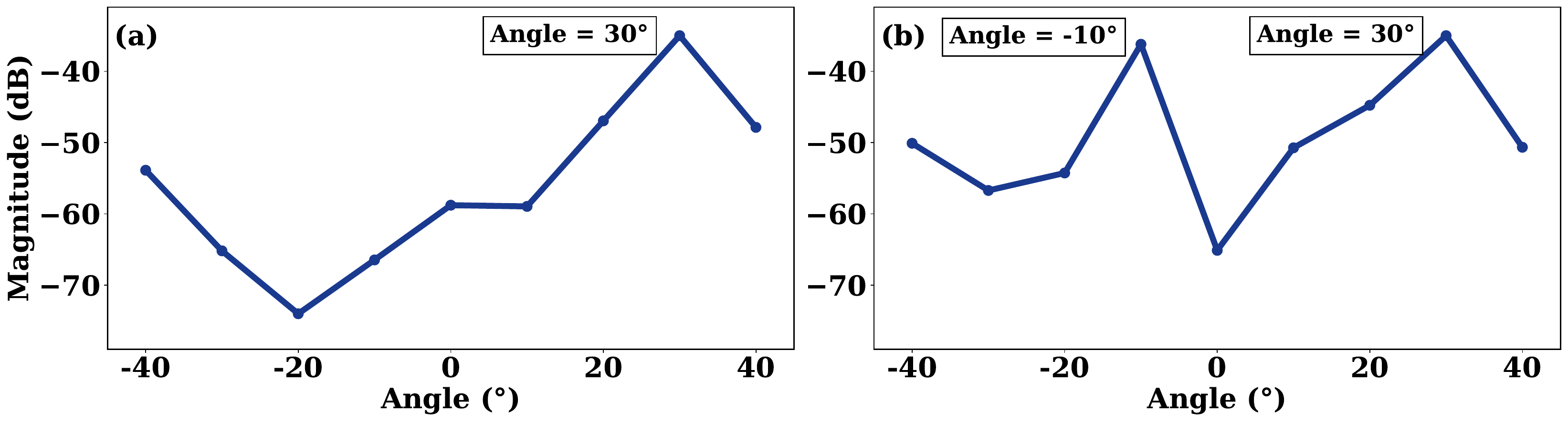}
    \caption{Azimuth angle comparison for (a) single target, and (b) multiple targets.}
    \label{fig:Angle_Output}
\end{figure}

\begin{figure}[!h]
    \centering
    \includegraphics[width=0.5\textwidth, trim=0cm 6cm 0cm 0cm, clip]{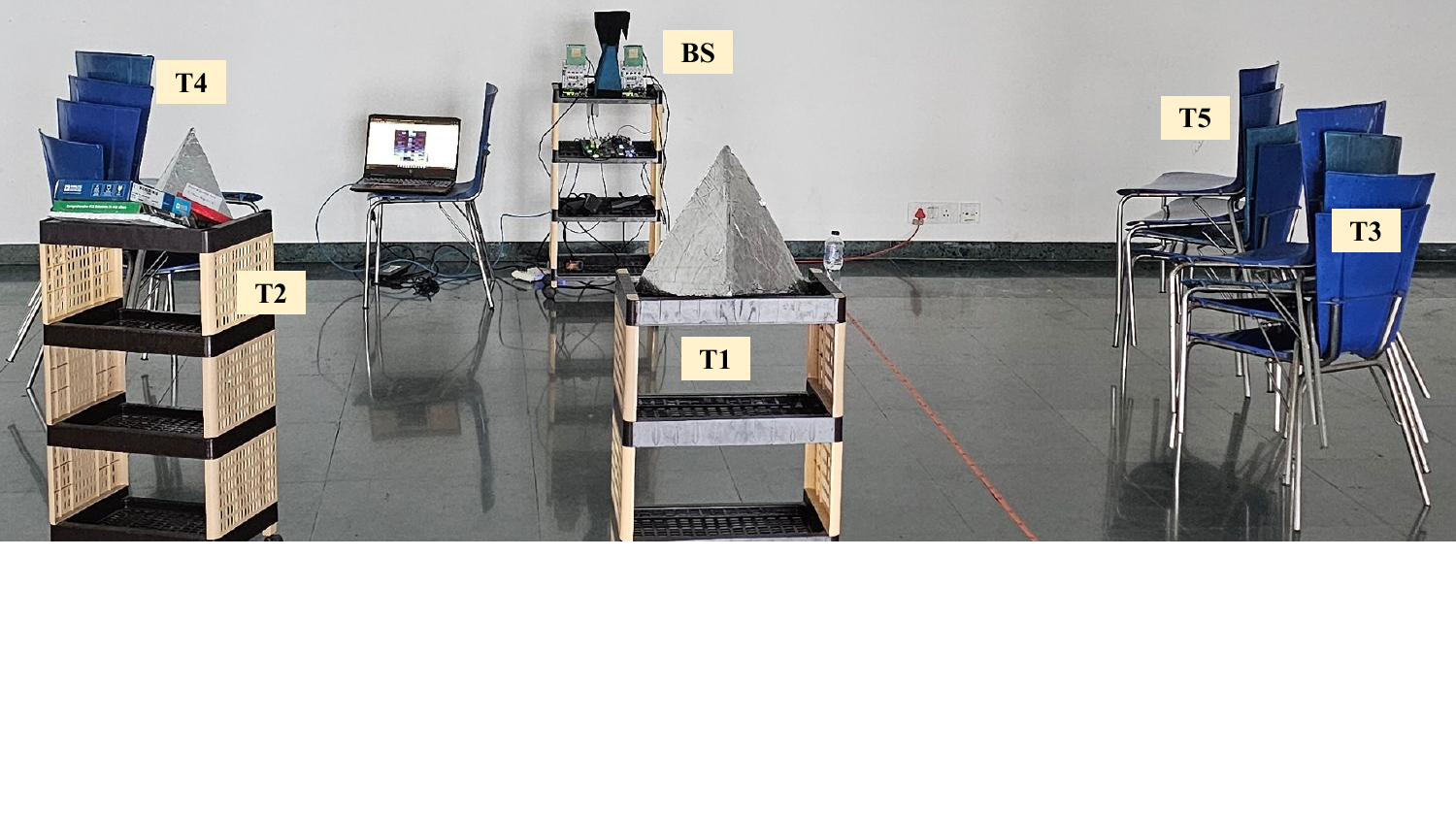}
    \caption{Experimental setup comprising five targets at different azimuths.}
    \label{fig:radar_exp_setup}
\end{figure}
\begin{figure*}[htbp]
    \centering
    \includegraphics[width=\textwidth,  trim=0cm 0cm 0cm 0cm, clip]{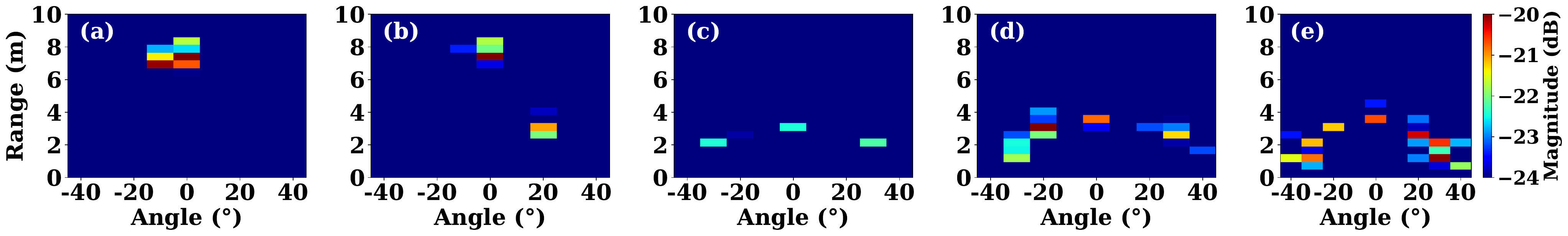}
    \caption{Performance of proposed 2D-radar architecture to resolve multiple targets in range-azimuth: (a) single target, (b) two targets, (c) three targets, (d) four targets, and (e) five targets for setup in Fig.~\ref{fig:radar_exp_setup}.}
    \label{fig:radar_tgt}
\end{figure*}

Next, the azimuth angle estimation performance is evaluated by placing a trihedral corner reflector at a fixed range of 3 m and varying the angular position. For each target position, beams are swept using analog beamforming over $\pm$40$^\circ$ in steps of 10$^\circ$, and the target's angular position is estimated from the peak. The experiment was conducted for both a single-target case, with a corner reflector at 27$^\circ$, and a two-target case, with corner reflectors at 27$^\circ$ and -8$^\circ$.

In the single-target case, the peak is detected at 30$^\circ$, while in the two-target case, the peaks are detected at 30$^\circ$ and -10$^\circ$, as shown in Fig.~\ref{fig:Angle_Output}(a) and (b) respectively. The estimation errors in both cases are bounded by the 10$^\circ$ angular-sweep step size. The experiment was then extended to 20 target positions spanning $\pm$40$^\circ$. The azimuth angle estimates benchmarked against ground truth are presented in Fig.~\ref{fig:RA_RMSE}(b), yielding an RMSE of 2.69$^\circ$.

Next, the performance of the ISAC transceiver in detecting multiple targets simultaneously is evaluated across five scenarios, with 1, 2, 3, 4, and 5 targets, placed at different ranges and azimuths as shown in Fig.~\ref{fig:radar_exp_setup}. The resulting range-azimuth maps obtained after radar signal processing are presented in Fig.~\ref{fig:radar_tgt}. The testbed successfully resolves all targets across all five scenarios. Table~\ref{tab:multi_target_error} compares the estimated angle and range of each target with the ground truth. The overall RMSE is $3.91^\circ$ in angle and $0.42$~m in range, and the angle error remains below the $10^\circ$ beam-steering step supported by analog beamforming. The detection performance is limited by angular resolution rather than range resolution, since the antenna aperture directly determines the angular resolution of the system. The radar system was unable to successfully resolve more than 5 targets in range and azimuth.
\begin{table}[t]
\centering
\caption{Target range and azimuth estimates for the single- and multi-target scenarios in Fig.~\ref{fig:radar_tgt}.}
\label{tab:multi_target_error}
\small
\setlength{\tabcolsep}{2.5pt}
{
\begin{tabular}{|c|c|c|c|c|c|}
\hline
\multirow{2}{*}{\textbf{Scenario}} & \multirow{2}{*}{\textbf{Target}} & \textbf{Actual} & \textbf{Estimated} & \textbf{Actual} & \textbf{Estimated} \\
 & & \textbf{Angle ($^\circ$)} & \textbf{Angle ($^\circ$)} & \textbf{Range (m)} & \textbf{Range (m)} \\
\hline
1 target & 1 & $-3$ & $0$ & $7.3$ & $7.0$ \\
\hline
\multirow{2}{*}{2 targets} & 1 & $-3$ & $0$ & $7.3$ & $8.0$ \\
\cline{2-6}
 & 2 & $17$ & $20$ & $2.4$ & $3.0$ \\
\hline
\multirow{3}{*}{3 targets} & 1 & $-28$ & $-30$ & $2.4$ & $2.0$ \\
\cline{2-6}
 & 2 & $-3$ & $0$ & $3.2$ & $3.0$ \\
\cline{2-6}
 & 3 & $28$ & $30$ & $2.4$ & $2.0$ \\
\hline
\multirow{4}{*}{4 targets} & 1 & $-34$ & $-30$ & $2.4$ & $2.0$ \\
\cline{2-6}
 & 2 & $-3$ & $0$ & $3.2$ & $3.5$ \\
\cline{2-6}
 & 3 & $15$ & $20$ & $2.4$ & $3.0$ \\
\cline{2-6}
 & 4 & $34$ & $40$ & $2.2$ & $2.5$ \\
\hline
\multirow{5}{*}{5 targets} & 1 & $-37$ & $-40$ & $0.8$ & $1.0$ \\
\cline{2-6}
 & 2 & $-14$ & $-20$ & $2.8$ & $3.5$ \\
\cline{2-6}
 & 3 & $-3$ & $0$ & $3.4$ & $3.5$ \\
\cline{2-6}
 & 4 & $14$ & $20$ & $2.8$ & $2.5$ \\
\cline{2-6}
 & 5 & $37$ & $40$ & $0.8$ & $1.0$ \\
\hline
\end{tabular}
\arrayrulecolor{black}
}
\end{table}
\subsection{Communication Performance Characterization}
The communication performance of the ISAC transceiver is evaluated in terms of EVM and BER.
\begin{figure}[!h]
    \centering
    \includegraphics[width=0.4\textwidth, trim= 0cm 0cm 4cm 0cm, clip]{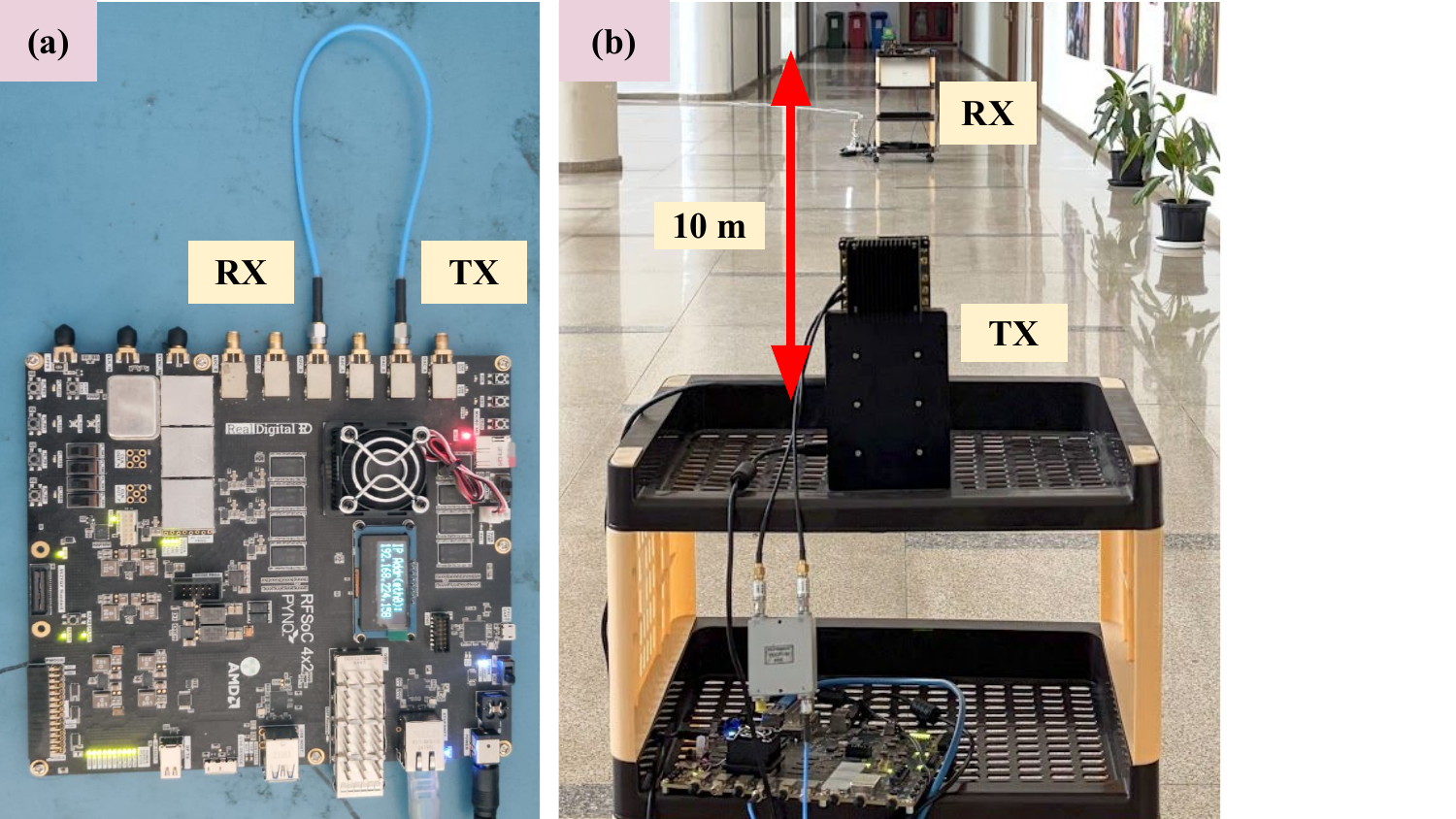}
    \caption{Experimental setup for communication in an indoor room environment: (a) loopback, (b) distance = 10 m (in corridor).}
    \label{fig:comm_near}
\end{figure}
The EVM is estimated as
\begin{equation}
\text{EVM} = \sqrt{\frac{\frac{1}{N} \sum_{n=1}^{N} \left| \mathbf{S}_{rx}[n] - \mathbf{S}_{tx}[n] \right|^2}{\frac{1}{N} \sum_{n=1}^{N} \left| \mathbf{S}_{tx}[n] \right|^2}}
\label{eq:evm}
\end{equation}
where $\mathbf{S}_{rx}[n]$ denotes the $n$-th received equalized complex symbol, and $\mathbf{S}_{tx}[n]$ represents the corresponding ground truth transmitted symbol. The BER is estimated as
\begin{equation}
\text{BER} = \frac{1}{N_{\text{total}}} \sum_{i=1}^{N_{\text{total}}} \mathbbm{1}\left[ \hat{b}_i \neq b_i \right]
\label{eq:ber}
\end{equation}
where $\hat{b}_i$ and $b_i$ denote the estimated and ground truth transmitted bits, respectively, and $\mathbbm{1}[\cdot]$ is the indicator function.

\begin{figure}[!h]
    \centering
    \includegraphics[width=0.45\textwidth,trim=0cm 0.2cm 0cm 0cm, clip]{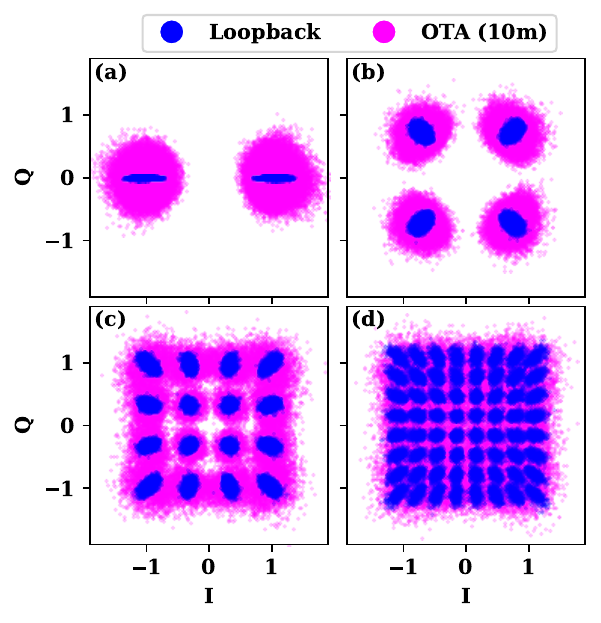}
    \caption{Constellation plots for wired loopback and OTA experiments for four modulation schemes: (a) BPSK, (b) QPSK, (c) 16 QAM, and (d) 64 QAM.}
    \label{fig:comm_const}
\end{figure}

\begin{figure}[htbp]
    \centering
    \includegraphics[width=0.5\textwidth]{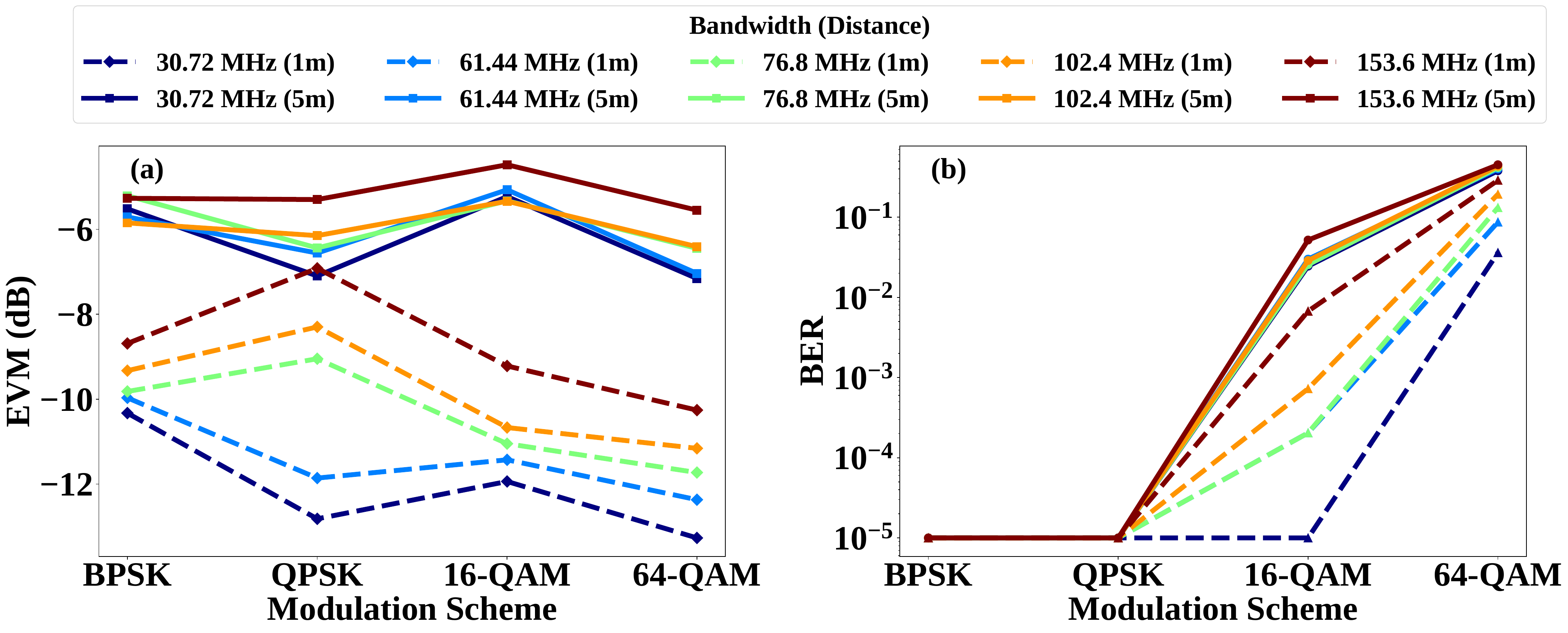}
    \caption{Performance of proposed communication architecture with MCS and bandwidth for different distances: (a) EVM, (b) BER.}
    \label{fig:comm_MCS}
\end{figure}

We consider the experimental setup shown in Fig.~\ref{fig:comm_near} for loop-back mode (a) and over-the-air mode with the BS-TX and MU-RX separated by 10 m (b). The received constellation points for four modulation schemes, namely BPSK, QPSK, 16-QAM, and 64-QAM, are shown in Fig.~\ref{fig:comm_const}. The received constellation points in the loopback mode, shown in blue, form tight clusters at the ideal symbol positions for all schemes, validating the functional correctness of the digital receiver chain. In the over-the-air case, the spread around the ideal constellation points is higher due to wireless channel fading.
Next, the EVM and BER of the ISAC transceiver are evaluated across different MCSs and presented in Fig.~\ref{fig:comm_MCS}(a) and (b), respectively. The EVM varies by only 2-3 dB across MCS because it depends primarily on the received SNR and hardware linearity, both of which are independent of the MCS. The BER, however, increases monotonically with modulation order due to the reduced noise margin between the more densely packed constellation points.
We now study the effects of bandwidth and distance on communication performance in terms of EVM and BER over a bandwidth of 30.72-153.6 MHz and a range of 1 m and 5 m. The corresponding hardware setup is the same as shown in Fig.~\ref{fig:comm_near}(b) with distance reduced to 1 and 5 m, in an indoor environment. While a wider bandwidth directly increases throughput, the integrated noise power at the receiver increases. As a result, both the EVM and BER degrade with wide bandwidths as shown in Fig.~\ref{fig:comm_MCS}(a) and (b). Similar degradation is observed with increasing distance due to higher path loss. During the performance evaluation, the radar bandwidth was varied from 153.6 MHz to 307.2 MHz to evaluate range resolution, whereas the communication bandwidth was limited to 153.6 MHz because the AFE output power decreases with bandwidth. Furthermore, the constant-envelope BPSK-modulated radar waveform permits near-saturation amplification, whereas the higher PAPR communication waveform requires power amplifier back-off, reducing the average transmit power.

\subsection{Joint Performance of Sensing and Communications}
The previous subsections characterized the sensing and communication subsystems independently. In this subsection, we evaluate how the interplay of radar and communication affects the other in a real radio environment.

\subsubsection{Effect of Beam Misalignment on Communication Performance}
\begin{figure}[htbp]
    \centering
    \includegraphics[width=0.5\textwidth, trim= 0cm 5.2cm 0cm 0cm, clip]{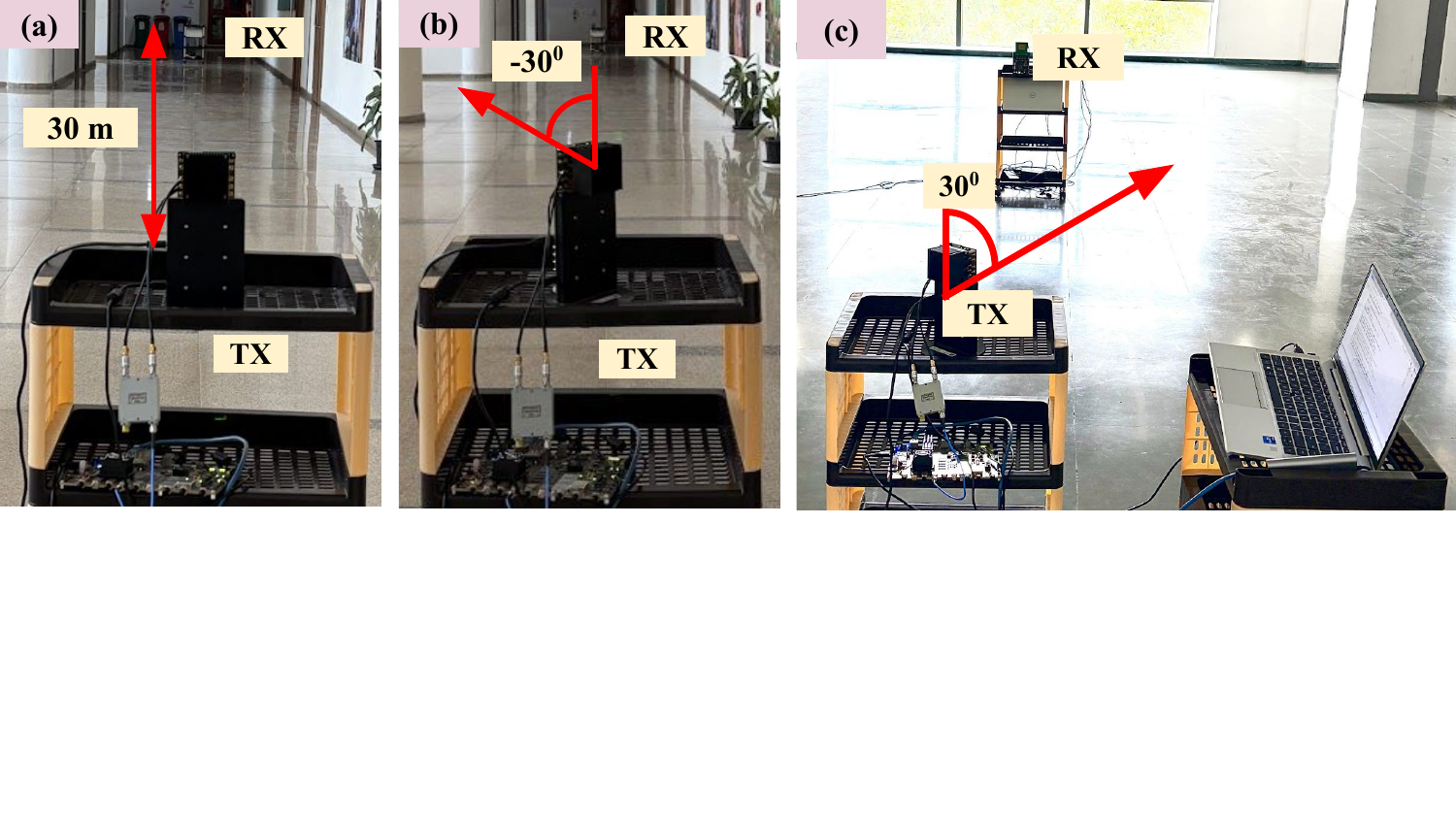}
    \caption{Experimental setup for communication system in corridor environment: (a) distance = 30 m, (b) beam misalignment = 30$^\circ$ (in corridor), and (c) beam misalignment = 30$^\circ$ (in indoor).}
    \label{fig:comm_far}
\end{figure}
\begin{figure}[htbp]
    \centering
    \includegraphics[width=0.5\textwidth]{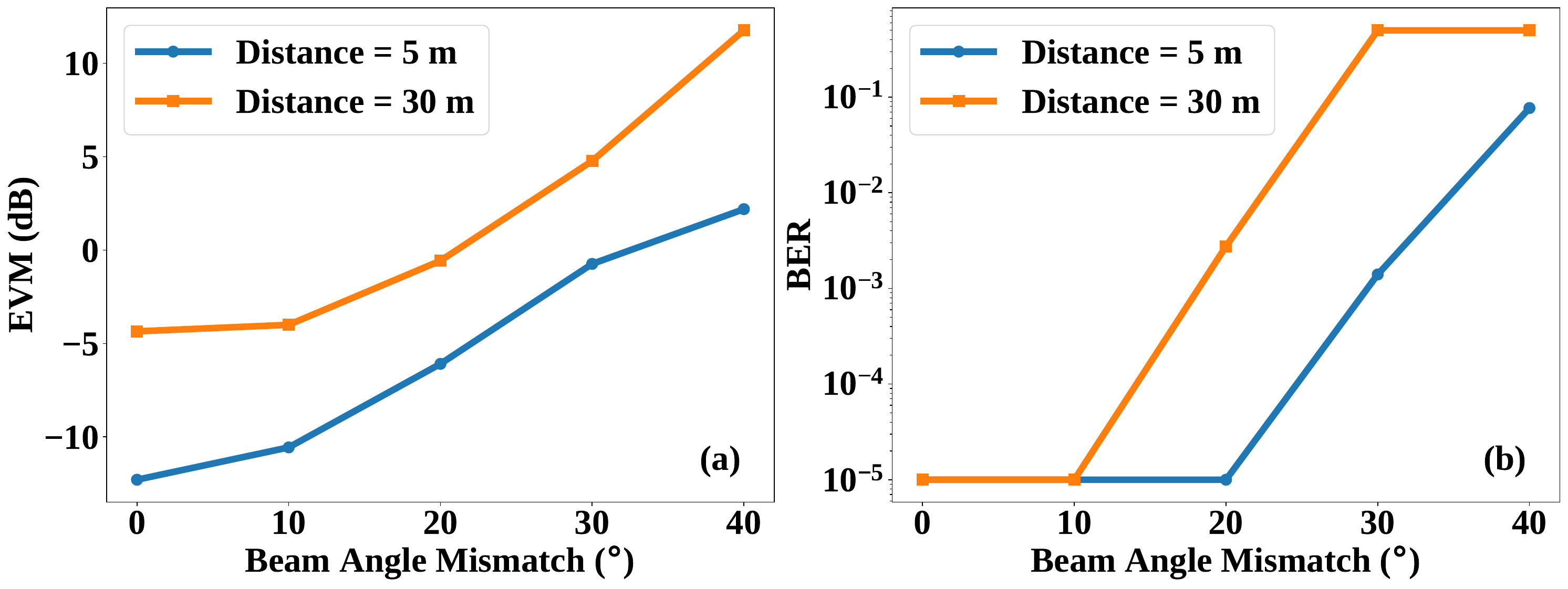}
    \caption{Performance of proposed communication architecture with angle and distance: (a) EVM, (b) BER.}
    \label{fig:comm_dist}
\end{figure}
\begin{figure}[htbp]
    \centering
    \includegraphics[width=0.5\textwidth]{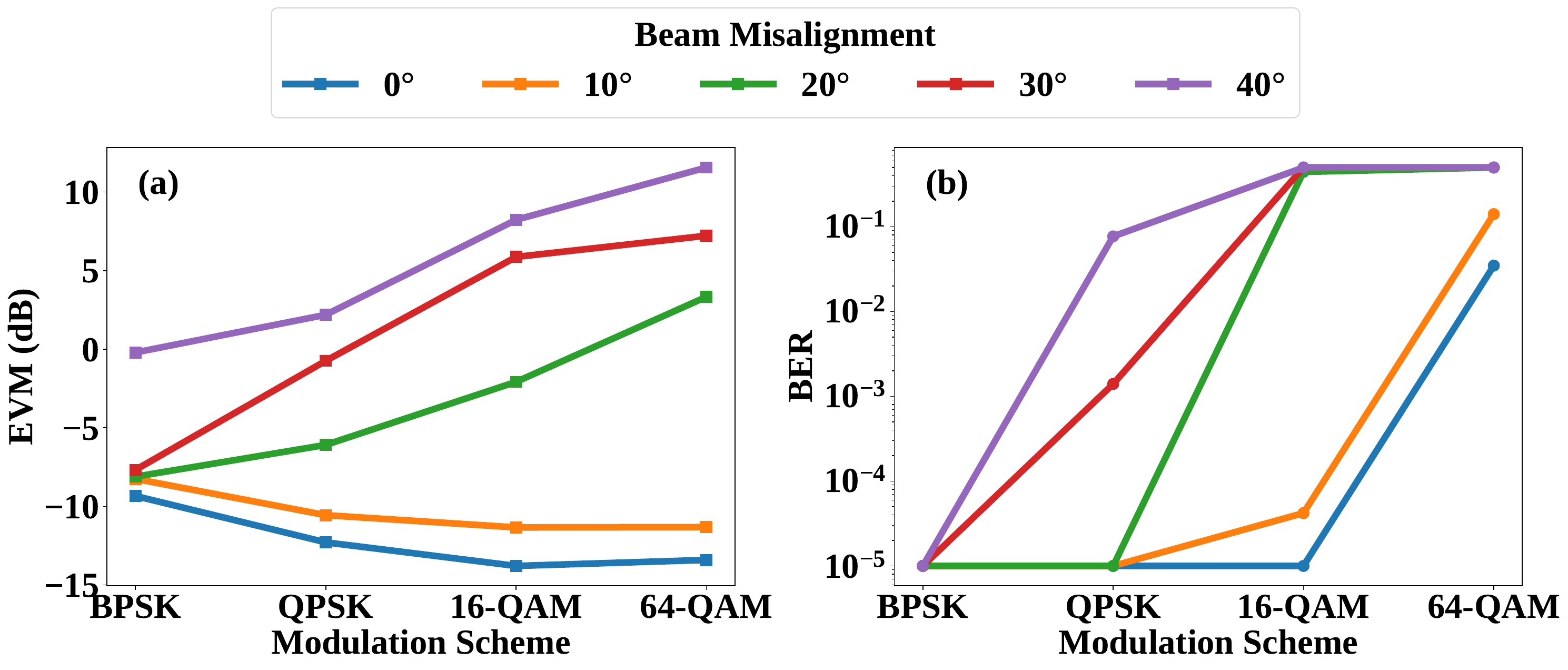}
    \caption{Performance of proposed ISAC architecture with angle mismatch: (a) EVM. (b) BER.}
    \label{fig:comm_angle_mismatch}
\end{figure}

Since the radar informs the communication functionality of the MU's azimuthal location, radar angular estimation errors directly affect the communication link quality through beam misalignment. We experimentally evaluate how the communication EVM and BER degrade due to radar beam alignment errors at two separation distances, 5 m and 30 m, with a 30.72 MHz bandwidth. The corresponding hardware testbed with 30 m separation between transmitter and receiver is shown in Fig.~\ref{fig:comm_far}(a). Further, the hardware testbed with 30$^\circ$ beam misalignment for 30 m and 5 m separation between the transmitter and receiver is shown in Fig.~\ref{fig:comm_far}(b) and (c) respectively. The results, presented in Fig.~\ref{fig:comm_dist}(a) and (b), show that both EVM and BER deteriorate with increasing angular mismatch, and the deterioration becomes more significant at larger distances due to increased path loss. At a distance of 30 m, a beam misalignment of 20$^\circ$ is sufficient to drive the BER above $10^{-3}$, leading to link outage.
Further, the effect of beam misalignment across different modulation schemes is evaluated at a fixed distance of 5 m and a bandwidth of 30.72 MHz. The EVM remains largely constant across modulation schemes at low beam misalignment, whereas the BER deteriorates with higher-order modulations due to reduced noise margins. With increasing misalignment, both EVM and BER worsen due to the resulting decrease in SNR, as shown in Fig.~\ref{fig:comm_angle_mismatch}(a) and (b) respectively. Even for robust modulations such as BPSK and QPSK, the BER degrades significantly when the misalignment exceeds 20$^\circ$, confirming that accurate beam alignment, facilitated by radar in ISAC, is critical for maintaining satisfactory communication link performance across all modulation orders.

\subsubsection{Real Time Target Tracking of Non-Static MU}
\begin{figure}[htbp]
    \centering
    \includegraphics[width=0.5\textwidth, trim= 0cm 2cm 0cm 0cm, clip]{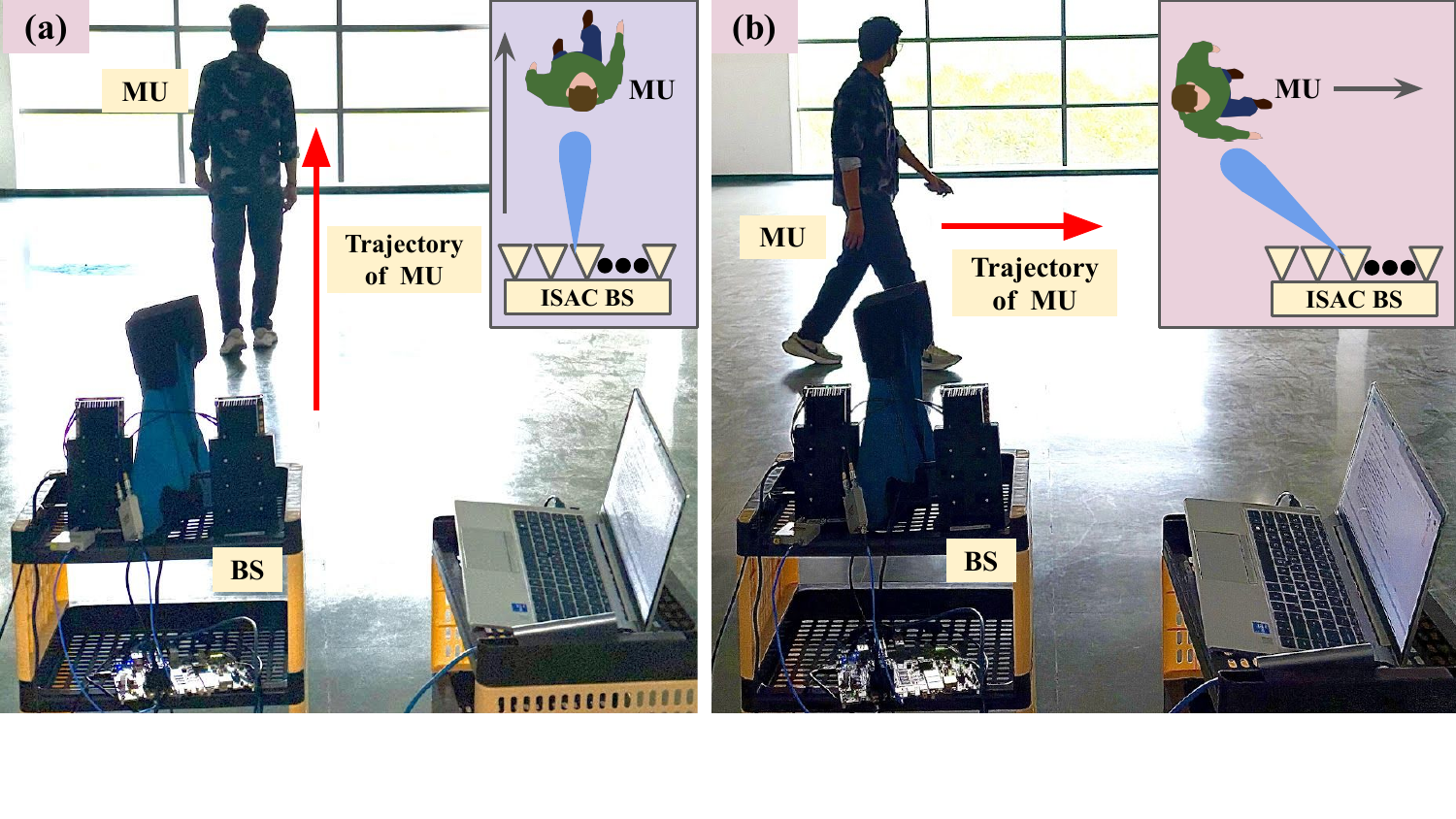}
    \caption{Target tracking experimental setup for (a) radial trajectory, (b) tangential trajectory.}
    \label{fig:Target_Traj}
\end{figure}
\begin{figure}[htbp]
    \centering
    \includegraphics[width=0.5\textwidth, trim= 0cm 5.8cm 0cm 0cm, clip]{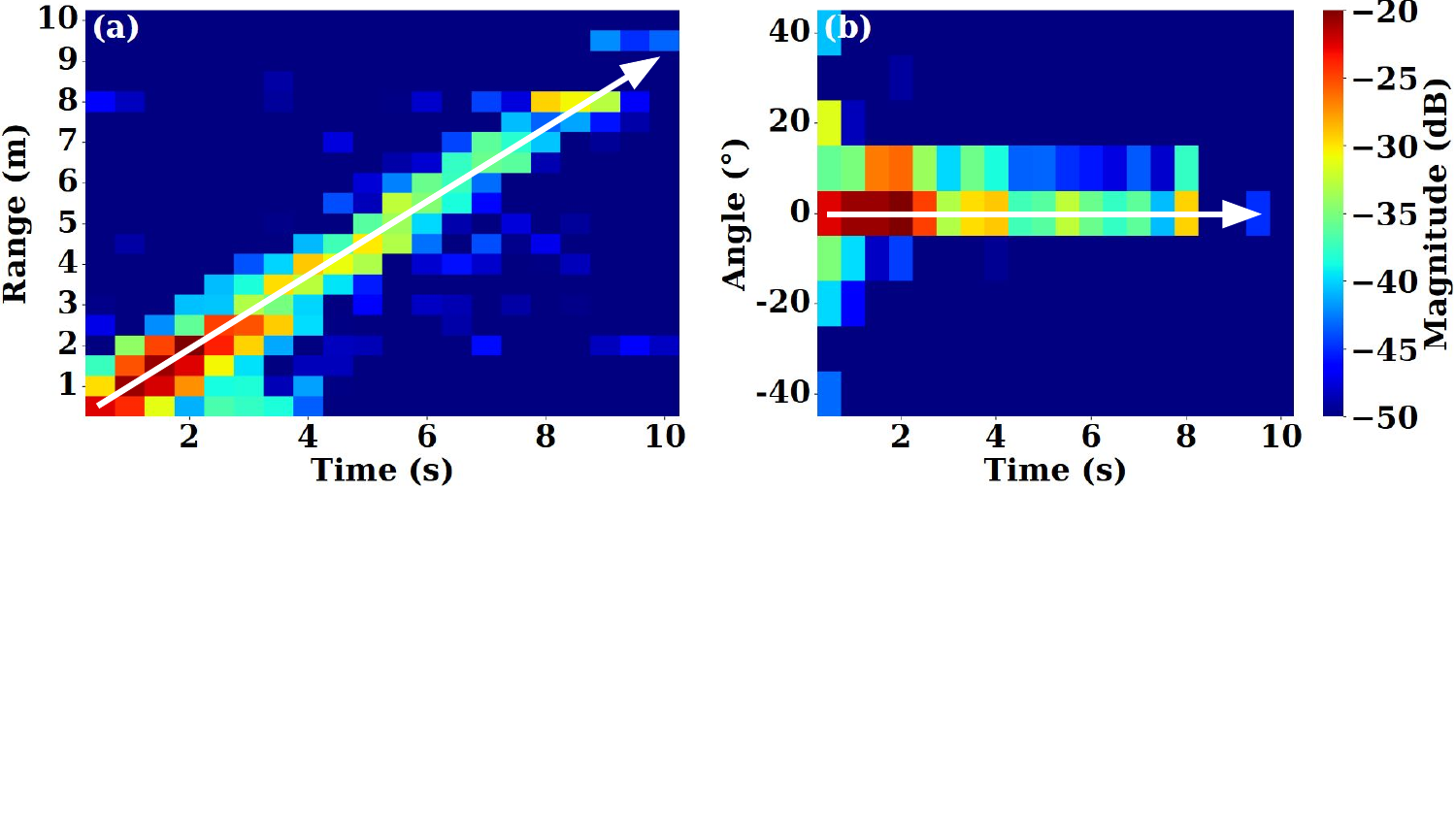}
    \caption{Target localization for radial trajectory: (a) Range, (b) Angle}
    \label{fig:radial_traj}
\end{figure}
\begin{figure}[htbp]
    \centering
    \includegraphics[width=0.5\textwidth, trim= 0cm 5.8cm 0cm 0cm, clip]{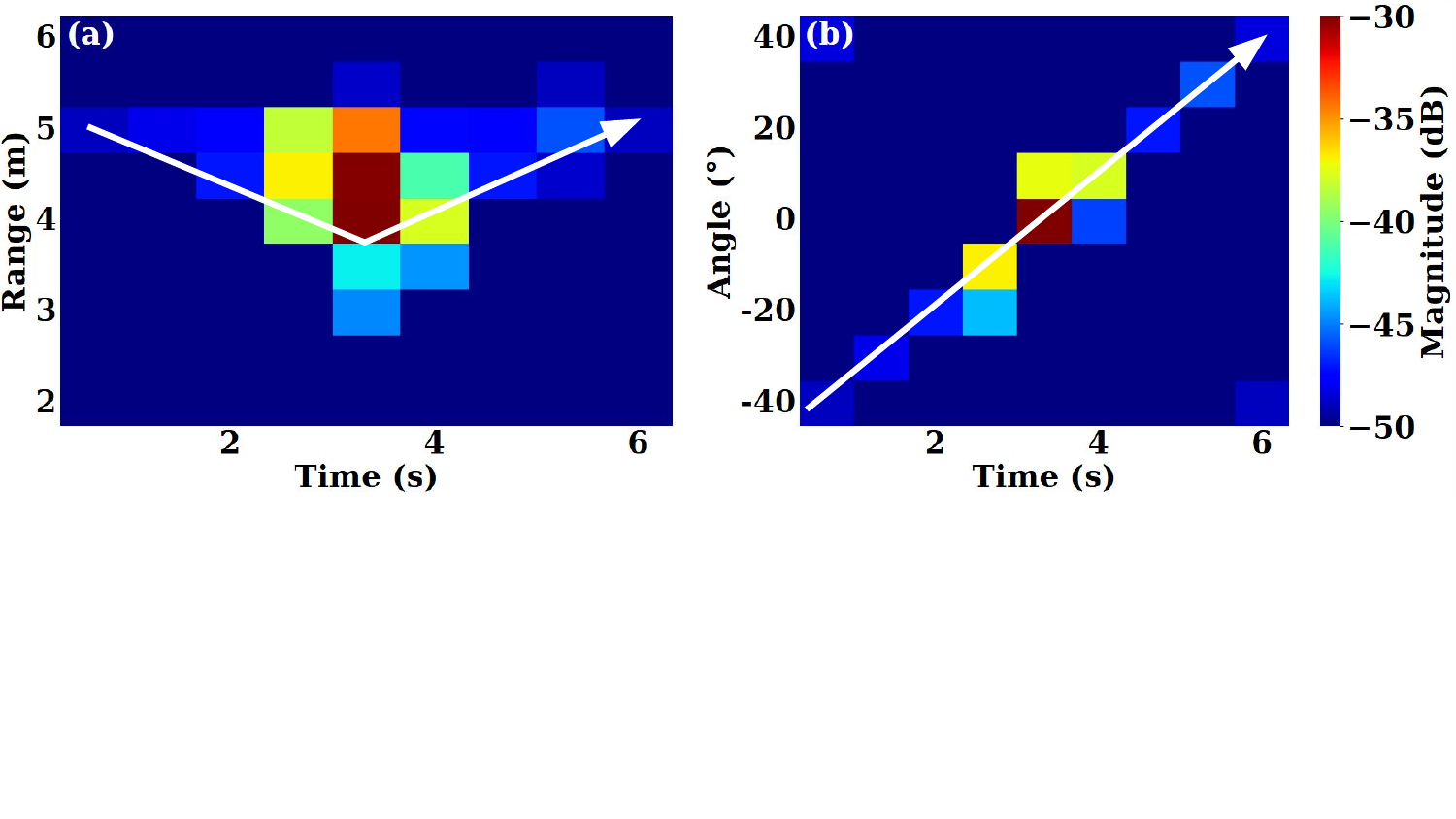}
    \caption{Target localization for tangential trajectory: (a) Range, (b) Angle}
    \label{fig:tangential_traj}
\end{figure}

Real-time tracking is essential for maintaining beam alignment. With the $0.5$~m range resolution, the $0.58$~s scan time limits the trackable velocity to $0.86$~m/s; hence, a walking human is used in these indoor experiments, which evaluate the tracking performance under multipath. To evaluate the tracking capability of the proposed testbed, a walking human is tracked along a radial trajectory for 10 seconds away from the ISAC BS, as shown in Fig.~\ref{fig:Target_Traj}(a). Similarly, we also consider a tangential trajectory for 7 seconds, as shown in Fig.~\ref{fig:Target_Traj}(b). The Doppler shift $f_m$, caused by motion of a target, introduces a phase rotation $\Delta\phi = 2\pi f_m T_{\mathrm{PRI}}$ between the two Golay sequences, which degrades their sidelobe cancellation. When the human moves along a radial trajectory with an average speed of $1$~m/s, the Doppler shift is $188$~Hz, and $\Delta\phi = 0.225^\circ$, which is negligible, and does not affect the autocorrelation properties of the radar waveform. Unlike the corner reflector experiments, a human is an extended target with multiple scattering centers and micro-Dopplers arising from the swinging motion of the arms and legs. A $1.74$~m tall human of $60$~kg is detected up to a range of $12$~m. The range, azimuth angle, and target strength are estimated from the peak of the range–azimuth map at every PRI of 3.33 $\mu$s. The resulting measurements are aggregated to construct range–time and angle–time plots, as shown in Fig.~\ref{fig:radial_traj}(a) and (b), respectively.
For the radial trajectory, the range increases linearly with time as the target moves away from the radar, as shown in Fig.~\ref{fig:radial_traj}(a). The strength of the returns falls as the target moves farther from the radar due to path loss, while the azimuth angle estimate remains constant near 0$^\circ$ with degrading amplitude, as shown in Fig.~\ref{fig:radial_traj}(b). For the tangential trajectory, the range first decreases as the target approaches and then increases as the target moves away, as shown in Fig.~\ref{fig:tangential_traj}(a). The strongest return occurs when the target is nearest to the radar at 3.4 s. The azimuth angle continuously increases from -40$^\circ$ to +40$^\circ$ over the duration of the trajectory, as shown in Fig.~\ref{fig:tangential_traj}(b). These results confirm that the radar successfully estimates the range and angle of a moving human target in real time, providing the necessary information for sensing-assisted beam alignment.

\subsubsection{Sensing-Assisted Beam Aligned Communication}

\begin{figure}[htbp]
    \centering
    \includegraphics[width=0.4\textwidth, trim= 0cm 0cm 7.2cm 0cm, clip]{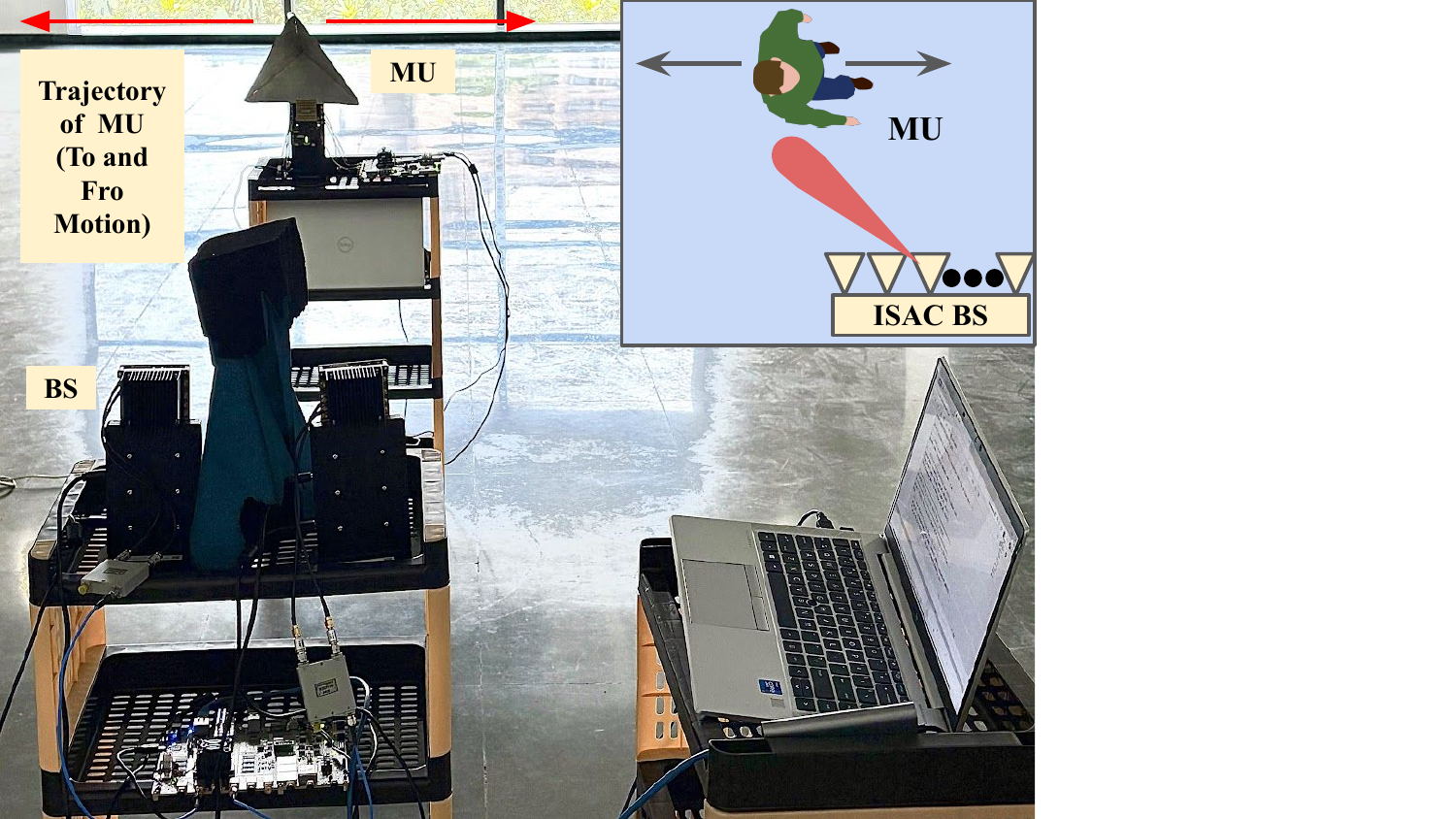}
    \caption{Experimental setup of the sensing-assisted beam-aligned communication using the proposed hardware testbed.}
    \label{fig:ISAC_MU}
\end{figure}

\begin{figure}[htbp]
    \centering
    \includegraphics[width=0.5\textwidth]{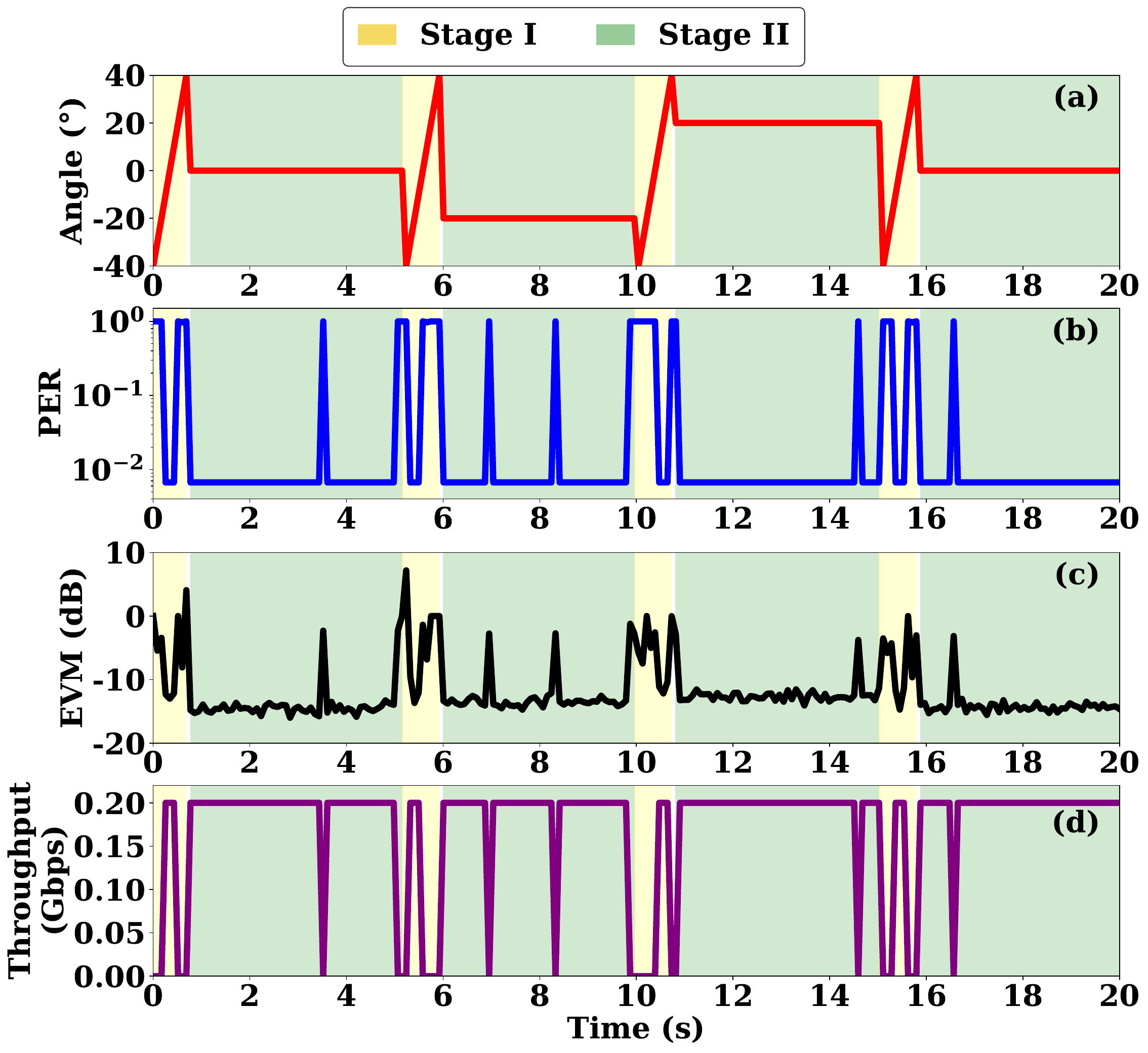}
    \caption{Performance of proposed ISAC architecture with moving target: (a) Angle, (b) PER, (c) EVM, and (d) Throughput}
    \label{fig:ISAC_BS_MU_EXP}
\end{figure}

The tracking results above demonstrate that the radar successfully estimates the range and angle of a moving target in real time. These estimates are now leveraged to steer the communication beam toward the MU. The full ISAC operation is demonstrated with MU moving along a tangential trajectory relative to the ISAC BS over 20 s, as shown in Fig.~\ref{fig:ISAC_MU}. The experiment is performed with a bandwidth of 153.6 MHz, a code rate of 0.5, and 16-QAM. The results are presented in Fig.~\ref{fig:ISAC_BS_MU_EXP}. The top row shows the radar-based angular sweep across 9 beams during Stage I and fixed beam transmission during Stage II. The second, third, and fourth rows show the corresponding packet error rate (PER), EVM, and throughput estimated from every 240 frames, respectively. The throughput of the communication subsystem is defined as:
\begin{equation}
\text{Throughput}
= f_s \log_2(M) R_c\left( \frac{N_d}{N_{\text{total}}} \right)(1 - \mathrm{BER})
\label{eq:throughput}
\end{equation}
where $f_s$ is the sampling rate, $M$ is the modulation order, $R_c$ is the code rate, $N_d$ is the number of data symbols per frame, and $N_{\text{total}}$ is the total number of symbols per frame including pilots and guard bands. For the system parameters of $f_s$ = 153.6 MHz, 16-QAM, $R_c$ = 0.5, and $N_d/N_{\text{total}}$ = 336/512, the peak throughput at near-zero BER is 0.20 Gbps. The factor $(1-\mathrm{BER})$ scales the throughput by the fraction of correctly received bits. Scaling it instead by the fraction of frames received without error, $(1-\mathrm{PER})$, where $\mathrm{PER}=1-(1-\mathrm{BER})^{N_b}$ and $N_b=N_d\log_2(M)R_c=672$~bits per frame, gives $\mathrm{PER}=0.67\%$ at the reported $\mathrm{BER}=10^{-5}$, and the throughput remains $0.20$~Gbps.

The ISAC BS initiates Stage I at the start of the experiment, during which the radar scans through 9 beams over a duration of 0.58 s to determine the location of the MU, as shown in Fig.~\ref{fig:ISAC_BS_MU_EXP}(a). During this interval, the communication link metrics are poor, and the throughput is low. Once the optimal beam is identified at 0$^\circ$, Stage II commences with improved PER, EVM, and a throughput of 0.20 Gbps. As the MU continues along the tangential trajectory, the beam becomes progressively misaligned, causing a fall in SNR. Stage I is re-initiated when the SNR falls by 3 dB, triggering a new beam search at 5.4 s. During this stage, the PER, EVM, and throughput again deteriorate. Once the optimal beam is identified, Stage II commences with restored communication metrics. This cycle repeats throughout the experiment, with the optimal beam identified at -20$^\circ$ at 6 s, at +20$^\circ$ at 11 s, and at 10$^\circ$ at 16 s. During Stage II, occasional degradation in communication metrics is observed due to multipath, hardware impairments, and noise, but the performance recovers without triggering Stage I, confirming that the degradation is not caused by beam misalignment. The throughput drops to zero during the Stage-I scans and at a few instants of Stage II, where the PER rises. Hence, the average throughput over the complete $20$~s experiment is $0.171$~Gbps, which is $85.5\%$ of the peak throughput in Stage II.

The beam alignment of the proposed work is compared with that of the sector-level sweep (SLS), which is the beam alignment procedure of the IEEE 802.11ad standard~\cite{IEEE80211_2016, SnehTVT2024}. In both methods, the MU sweeps its $9$ beams and recovers one packet in each beam. In the SLS, the MU also correlates the beam refinement fields of all $81$ beam pairs and sends the best beam of the BS back through an uplink packet, which requires $113.28$~ms. In the proposed work, the BS finds its beam through the radar scan, which requires $44.82$~ms. The beam switching time is the same in both methods and is excluded from this comparison. The calculated beam alignment latency of the proposed work is $40\%$ that of the SLS.

\section{Summary}
\label{sec:conclusion}
In this paper, we have presented the design and implementation of a fully integrated end-to-end mmWave ISAC testbed operating in the FR2 frequency band. The DFE is implemented using hardware software co-design on the Xilinx RFSoC 4x2 platform, while the AFEs for the BS and MU are implemented using Sivers Semiconductors EVK02004 and EVK02001, respectively. The testbed is experimentally validated in real radio environments across single and multiple target scenarios. The sensing subsystem achieves a range resolution of 0.5 m, an azimuth resolution of $10^\circ$, and an angular sweep duration of 0.58 s using analog beamforming. The testbed successfully resolves up to 5 targets simultaneously in the 2D range-azimuth space. High sensing accuracy is achieved by implementing a custom IP core in the programmable logic, which mitigates non-deterministic pipeline jitter inherent to the RFSoC platform. The communication subsystem is characterized across different MCSs at distances up to 30 m, with performance evaluated in terms of BER, EVM, and throughput. We further evaluate the impact of radar beam misalignment on communication link performance, where higher-order MCSs are found to be particularly sensitive to beam misalignment errors. Finally, we demonstrate radar-assisted beam-aligned communication between a fixed BS and an MU moving along a tangential trajectory. In future work, we will extend the testbed to support Doppler velocity estimation for improved tracking and experimentally study the relation between the radar and the communication channel.

\section*{Acknowledgment}
This work is supported by project funding from the Ministry of Electronics and Information Technology (MeitY), the TCS PhD Fellowship (2024–2028), the Qualcomm Innovation Fellowship (2024), the Chips to Startup (C2S) Programme, the ANRF-ARG grant, and the DST-PURSE grant (SR/PURSE/2023/172), Govt. of India.

\bibliographystyle{ieeetr}
\bibliography{ref}

\end{document}